\documentclass[default,iicol]{sn-jnl}
\usepackage{natbib}
\usepackage{amsmath,amssymb,amsthm,csquotes}

\setcitestyle{numbers,open={[},close={]}} 

\begin{document}

\title[Article Title]{An NLP-Assisted Bayesian Time Series Analysis for Prevalence of Twitter Cyberbullying During the COVID-19 Pandemic}

\author[1]{Christopher Perez (cperez7@ufl.edu)}
\author[1]{Sayar Karmakar (sayarkarmakar@ufl.edu)*}
\affil[1]{\orgdiv{Department of Statistics}, \orgname{University of Florida}, \orgaddress{\city{Gainesville}, \postcode{32601}, \state{FL}, \country{USA}}}
\affil*[]{\orgdiv{Corresponding Author}}

\begin{sloppypar}

\abstract{COVID-19 has brought about many changes in social dynamics. Stay-at-home orders and disruptions in school teaching can influence bullying behavior in-person and online, both of which leading to negative outcomes in victims. To study cyberbullying specifically, 1 million tweets containing keywords associated with abuse were collected from the beginning of 2019 to the end of 2021 with the Twitter API search endpoint. A natural language processing model pre-trained on a Twitter corpus generated probabilities for the tweets being offensive and hateful. To overcome limitations of sampling, data was also collected using the count endpoint. The fraction of tweets from a given daily sample marked as abusive is multiplied to the number reported by the count endpoint. Once these adjusted counts are assembled, a Bayesian autoregressive Poisson model allows one to study the mean trend and lag functions of the data and how they vary over time. The results reveal strong weekly and yearly seasonality in hateful speech but with slight differences across years that may be attributed to COVID-19.}


\maketitle

\section{Introduction}
Technological developments throughout history have fundamentally changed how people communicate and interact with one another. With new successes come new challenges, as the rapid proliferation of the internet has led to a phenomenon known as cyberbullying. Many questions may arise, such as how cyberbullying can propagate and how it interacts with global crises such as the COVID-19 pandemic. In this paper, we combine natural language processing and Bayesian time series analysis methods to provide a systematic assessment of the trends of cyberbullying for a span of three years. We begin this exposition with an informal description of cyberbullying before describing the connection with COVID-19 and Twitter. 

\subsection{Describing Cyberbullying}

 One definition for cyberbullying is "\emph{An aggressive, intentional act carried out by a group or individual, using electronic forms of contact, repeatedly and over time against a victim who cannot easily defend him or herself}" \citep{smith2008cyberbullying}. And with increases in computer use, the possibility for cyberbullying grows. 

An immediate question is what sets cyberbullying and traditional in-person bullying apart. These differences are paramount when considering a theoretical approach to studying cyberbullying, as models made with traditional bullying in mind may not tell the two apart \citep{barlett2017theory}. One major point is that cyberbullies can maintain anonymity online which makes it difficult to locate perpetrators \citep{bonanno2013cyber}. Due to the reach of social networks, cyberbullying may also persist far beyond the reaches of normal bullying and can proliferate to large swaths of people, often attaining viral status \citep{aboujaoude2015cyberbullying}. Cyberbullying is a perpetual phenomenon that constantly places stress on the victim.

Cyberbullying has been studied to be a cause of many negative outcomes in victims. A meta-analysis conducted in the topic reveals correlations with low self-esteem, depression, and drug abuse \citep{kowalski2014bullying}. Many episodes of attempted suicide and self-harm have been directly attributed to cyberbullying \citep{kwan2020cyberbullying}. Interestingly, according to surveys, only a small percentage of cyberbullying victims are not bullied in a traditional, in-person manner \citep{olweus2018some}. Regardless, the many negative outcomes and pervasiveness of cyberbullying has led to some to suggest it is a serious public health threat, and its danger can only grow with increased mobile device and social media usage \citep{aboujaoude2015cyberbullying}.

\subsection{COVID-19 and Cyberbullying}
The proliferation of COVID-19 has significantly changed the way many people live. Lockdowns have been put in place to curb the spread, but not without consequences. As humans thrive in social situations, the isolation of many from the day-to-day affairs has led some to posit that people's mental health will worsen, noting consequences such as maladaptive behaviors, loneliness, and depression \citep{talevi2020mental}. Direct research shows that quarantines and self-isolation are linked with higher prevalence of issues like depression and insomnia \citep{wang2021impact}. When comparing time periods before and after the COVID-19 pandemic, \citet{barlett2021comparing} suggest that important components of their cyberbullying model, such as cyberbullying attitude, cyberbullying behavior, and belief in irrelevance of muscularity in online bullying, have significantly changed between these time points. It is also observed that cyberbullying is correlated with COVID-19 experiences \citep{barlett2021cyberbullying}. 

The interaction of COVID-19 and cyberbullying in academic settings is a topic of great interest. As universities and schools around the world shifted to online instruction to deter spread of the virus, approximately 1.5 billion students have had their education interrupted \citep{bozkurt2020global}. The effects on isolation on university students is an important matter, as these groups show high proportions of common mental disorders. Such literature reveals these groups were associated with more frequent internet use and may thus have increased probability of being involved in cyberbullying \citep{mota2021mental}. Quantitative analysis of such groups in India \citep{jain2020has} have shown that 80\% of those between 17 and 18 years old were bullied during the pandemic, and 79\% of those experiencing traditional bullying before the pandemic were cyberbullied during pandemic, corroborating the notion that victims of traditional bullying are also victims of cyberbullying. Other categories of victims show an increase in percentage of those cyberbullied from before to during the pandemic as well.

\subsection{Using Twitter to Understand Social Phenomena and Cyberbullying}
As in the aforementioned studies, much research in cyberbullying involves the use of survey data. However, if one relies on a responder's willingness to self-report, then that leaves the door open to problems such as responder bias and invalid responders. In a study on adolescent regarding risk behavior, it is found that responders who purposefully answer wrongfully "showed" higher rates of such behavior, such as alcohol and drug consumption \citep{cornell2012effects}. Such a happening may be present in bullying surveys as well, which can be exacerbated by sample-size limitations imposed by cost and time. Additionally, if a researcher was interested in studying the impact of COVID-19 on cyberbullying, they ideally have to collect data before the pandemic began, as done in previous research \citep{barlett2021comparing}. Since the magnitude and impact of global crises are sometimes unforeseeable, it can be difficult to know when to start such a longitudinal study.

Social media can serve as rich data source that remedies some of the issues that affect survey collection. One major advantage is the scope of social media. Twitter, the choice for our study, had approximately 186 million users, 36 million of which from the United States~\footnote{\url{https://www.businessofapps.com/data/twitter-statistics/}} in 2020. Furthermore, the availability of web scraping technology and even an official API allow individuals to use Twitter's public archive of tweets dating back to 2006. This software is often free to use, making large-scale studies much more affordable. Information harvested from these tweets can be used for many purposes, such as monitoring disease spread and forecasting elections \citep{signorini2011use, tumasjan2010predicting}. This data source enables us to perform a study reminiscent of longitudinal study, with the ability to collect previous years of data without the associated cost of maintaining a large-scale study for many years.

However, one may question the efficacy of using social media such as Twitter as a data source to understand social phenomena. For example, \citet{tumasjan2010predicting} showed that using Twitter traffic to predict vote share in German errors resulted in very low prediction error. \citet{signorini2011use} demonstrate a correspondence between Twitter data and the H1N1 at the overall national level, as well as smaller geographic regions. Based on real-time data, their estimates could be produced earlier than regular health reports. While these present advantages, one must also consider the possible limitations. For instance, predicting elections using Twitter faces two major issues, one being that sampling data from social media does not match the sophistication of more developed polling processes, and that spam, propagandists, and fake accounts can easily manipulate data \citep{gayo2011limits}. In the influenza study conducted by \citet{signorini2011use}, the researchers faced limitations in the lack of uniformity of Twitter usage by different locales and in different time periods. They also could not generalize to a population beyond some form of Twitter population. These limitations are also present in the current study. Brief use of the search endpoint for an arbitrary query may show a few spam or bot posts, heightening the importance of filtering these posts out. Additionally, since Twitter's sampling algorithm is unknown, it is difficult to identify the exact population the study's results can be generalized to. There are methods to reduce this uncertainty by using geo-tagged tweets, as retrieving tweets of this nature results in a more complete sample \citep{morstatter2013sample}.

Despite limitations, there are a variety of points to make in justifying Twitter as a data source to study cyberbullying. \citet{mchugh2019omg} suggests that Twitter is a hotbed for "intentionally aggressive, harmful communication." They explain results from surveys showing that about 70\% of college students use Twitter, and the amount of cyberbullying on Twitter was gauged to be higher than other platforms such as Facebook and Instagram. Additionally, the official Twitter API~\footnote{\url{https://developer.twitter.com/en/docs/twitter-api}} allows users to perform a variety of queries and searches on its public archive. With this technology, we can collect swaths of tweets satisfying certain specifications, such as containing particular keywords or hashtags, in order to study the cyberbullying problem.

To analyze the tweets retrieved from Twitter in the first place, many researchers turn to natural language processing (NLP). The usage of NLP in this study is similar to that of previous work done regarding Twitter cyberbullying and COVID-19 \citep{babvey2021using}. The study by \citet{babvey2021using} uses NLP as a pre-processing step to filter out tweets that have low probability of being abusive speech, and then examine the difference in number of such tweets before and after a fixed time period. In this study, we use NLP to perform a similar pre-processing step, but instead of comparing a fixed time point, we study each day from 2019 to 2021. Few studies have attempted to study cyberbullying in response to COVID-19 from a continuous perspective. Researchers using Google Trends time series data \citep{bacher2022covid} demonstrate that cyberbullying was actually disrupted by COVID-19, while others, using Twitter data \citep{karmakar2020evaluating}, claim that it led to an increase. We seek to address these studies by broadening the time frame of analysis and using more thorough methods, namely NLP, to obtain better samples to study cyberbullying patterns with.

\subsection{Considerations and Assumptions for Cyberbullying Detection} 
As we seek to understand the general volume of cyberbullying over several years of time, we are forced to make certain assumptions that allow us to work with the appropriate type of data. In light of our choice of a pre-trained NLP model, we assume that two metrics, hatefulness and offensiveness of a tweet, are associated with cyberbullying. However, detecting cyberbullying in an extremely accurate fashion may be a more difficult task than this assumption may imply. Many detection models in literature employ more advanced considerations to detect cyberbullying events such as images, location, and a given user's profile and comment history, which may contain vital information to predict cyberbullying behavior \citet{cheng2019xbully,dadvar2013improving}. Other models involve hierarchical attention networks to make use of the inherent structure of social media as additional context for predicting \citep{cheng2019hierarchical}, or constructing graphs composed of sender and receiver nodes to mimic cyberbullying interactions \citep{huang2014cyber}. In our study, the NLP model of choice does not necessarily make use of this more complicated information, and thus its accuracy may suffer relative to more state-of-the-art methods. Regardless, we choose this model for its off-the-shelf accessibility and hence tailor our analysis to what the model is capable of delivering.

\subsection{Summarizing Our Contribution} 
For our study, we collected 1,004,466 tweets from January 1st, 2019, to December 31st, 2021, with the Twitter API's search endpoint based on keywords used in previous cyberbullying and abusive speech studies \citep{wiegand2019inducing, nand2016bullying, cortis2015analysis}. To clean the data, a pre-trained NLP model tuned to classification of offensive and hateful tweets models the probability of a given tweet being offensive or hateful \cite{barbieri2020tweeteval}. Because of irregularities in the search endpoint's returned sample sizes, the count endpoint is used with the same keywords as before to obtain a more consistent and comprehensive count of all tweets that the search API could have picked from. The tweets from the search endpoint are then filtered using certain probability thresholds to select relevant tweets. Then, for each day in the study, the fraction of relevant tweets out its respective sample is calculated. This fraction is multiplied by the number reported by the count endpoint, generating a time series that takes into account the different proportions of abusive content on each day while addressing issues with the search endpoint's sampling procedure. This data collection and filtering is described in section 3.

Visual analysis of the data collected is in section 4 and motivates our usage of a Bayesian time series model in section 5. The results demonstrate that the tweets likely to be hateful speech exhibit strong weekly and year seasonality, which is not as evident in the unfiltered data. Patterns such as two distinct increases in mean trend during the first and second halves of the year remain constant throughout 2019 through 2021, but with slight differences. A time series of new COVID-19 cases is fit to the same model and similarities between this model and the Twitter data are discussed. Further, the effect of the pandemic on these trends, if at all, is to decrease the scale of potential cyberbullying tweets, as well as widen the second peak of the year. The proportion of these potential cyberbullying tweets from their respective daily samples also seems to roughly decrease over time. The results can help guide developers of Twitter or other social media to ramp up mitigation technology at the appropriate time by taking the strong seasonal behavior into account, and overall, introduce novel ways of examining a familiar problem.

The work is concluded by a discussion of the issues and limitations of the study, the implications of the results found, and avenues for future research.

\section{Related Work on Internet Data, Cyberbullying, and COVID-19}
This study is motivated by the findings of \citet{karmakar2020evaluating}, which employed a Bayesian, time-varying linear Poisson autoregressive model to tweet counts containing keywords related to cyberbullying. Such an analysis was the first of its kind in this field. Their study, confined to the first half of 2020, concluded a rise in mean trend from March to April similar to that of COVID-19 cases and that the first lag accounted for most of the correlation. However, there is criticism to be made in that preliminary analysis. The choice of keywords along with lack of text analysis confined the results to cyberbullying discourse rather than actual cyberbullying events. Cyberbullying attacks may precipitate awareness and discourse, but they do not follow the same time series. While their work collected data using web scraping, the current study pulls data directly from the Twitter API.

Another work by \citet{bacher2022covid}, instead of using Twitter data, opts to use Google Trends, a site that provides time series data for search intensities of search terms~\footnote{\url{https://trends.google.com/}}. This work studies search intensity of cyberbullying and bullying. Like the previous study, they do not depict the frequency of cyberbullying events exactly. Further, since the main model in our study is based on a Poisson distribution, which requires count data, we cannot use the TVBARC model on Google Trends data in attempt to replicate the study.

However, consider the two following similar findings from these works. The model fit by \citet{karmakar2020evaluating} reveals an increase in mean trend from around March to May of 2020, while the model constructed by \citet{bacher2022covid} shows that the deviation from predicted log search intensity increases roughly around the same time frame. One important distinction, though, is that the Bacher-Hicks study includes data from before January 2020 and slightly after. Bacher-Hicks claims, further, that this rise in the March-May period was just an increase back to levels before the onset of the pandemic, and that COVID-19 had disrupted cyberbullying. In the absence of a wider time frame in Karmakar's paper, one may conclude an increase in cyberbullying discourse on Twitter, but this may very well suggest a return to pre-pandemic levels as Bacher-Hicks describes. Again, one cannot say that this necessarily extends to cyberbullying events, but how it becomes a trending topic over time. Furthermore, social media and search engines are used with different motives in mind, which may result in discrepancies in findings \citep{li2021comparing}.

The work done by \citet{babvey2021using} motivates our decision to employ an NLP model in an attempt to retain true cyberbullying events. They query Twitter for keywords associated with abusive speech and then run a machine learning model to discard tweets that are likely not abusive. By using such methods, they are able to have more confidence that their data can represent actual cyberbullying events. They compare two sets of data collected before and after March 2020 to gauge the effect of COVID-19 and the associated interactions with cyberbullying. Their results show an increase in prevalence of abusive and hateful tweets once the pandemic-era lockdowns began.

Usage of NLP methods can help avoid the issue of covering cyberbullying discourse rather than potential cyberbullying events. The use of a larger time frame along with a more continuous time series approach allows one to see whether COVID-19 has a sustained effect on Twitter cyberbullying, or if previous findings may have been coincidences or one-time occurrences. Now, the data collection and NLP filtering procedure are discussed.

\section{Data Collection and Cleaning}
A straight-forward way to access to Twitter data is by using the official Twitter API. To make the most use of the Twitter API, we were provided with an Academic Research License, granting us features such as the full-archive search. This is critical to our research, as it lets us search many years' worth of data quite easily. We used the R programming language to interact with the API.

\subsection{A Foreword on the Twitter API}

Unfortunately, the algorithm used by the Twitter API to sample tweets is unknown. As studied by \citet{thelwall2015evaluating}, the search endpoint may not be comprehensive. However, the tweets that were not retrieved by its sampling procedure are more likely to be spam. Other works \citep{morstatter2013sample} point out that the sampling tweets may result in decreased accuracy (as compared to alternative, costly methods to acquire every single tweet), but interestingly, the sampling algorithm recovers a higher proportion of tweets that are geo-tagged.

The nature of time series analysis emphasizes these sampling issues. The subset of tweets taken from all matching tweets may not be a fixed percent, so the relative sizes between daily counts is not preserved. Simple repetitions of identical requests may sometimes return more tweets seemingly at random. Comparison with the number reported by the count endpoint is an enticing option for a couple reasons. The count endpoint returns much more consistent results through runs, and since it does not have to go through additional compliance that the search endpoint does, the data returned may be more complete~\footnote{\url{https://developer.twitter.com/en/docs/twitter-api/tweets/counts/introduction}}. To this end, we take the percentage of relevant tweets from a given daily sample and multiply that percentage to the number reported by the count endpoint. Selection of relevant tweets is described below.

\subsection{Collection Procedure}
To collect a representative dataset, we first sample tweets using the search endpoint to access their textual content. A list of keywords must be assembled to query for in both the search and count endpoints. As a starting point, we reference a lexicon provided by \citet{wiegand2019inducing}. It contains a list of words each with a score rating its abusiveness according to their trained model. One may notice that identical words appear more than once (as a verb and as a noun, for instance), and hence we only keep the highest score and discard the other entries. From this adjusted list, the 100 highest ranking words were taken.

For comparison purposes, we reference two similar studies of cyberbullying analysis \citep{nand2016bullying, cortis2015analysis}. Their lists of keywords are different and are only composed of about 25 and 10 words respectively. To test the efficacy of our 100 keywords, we sample tweets in January 2020 using our original list of words, specifying no retweets, written in English, and based in the United States. We then count the number of tweets each keyword in our list appeared in, including keywords in the other studies that were not in our original list. Based on numbers of tweet occurrences, we again take the 100 highest performing words.

With this new list that combines information from the aforementioned studies, we use the search endpoint to query the entirety of 2019, 2020, and 2021. Like before, we specify no retweets, tweets written in English, and tweets from the US, but this time we also specify no promotional tweets. Each day contains anywhere between 400 and a couple thousand tweets with their textual content. In total, we collected 1,004,466 tweets. The same query is also used for the count endpoint to collect daily counts. Once the full data set is assembled into a data frame, the results are written into CSV files for storage.

\subsection{A Pre-trained NLP Model for Pre-processing Data}
As discussed before, in order to assemble a time series of cyberbullying events, certain assumptions may be made. Using user profile information in a large-scale time series context may prove to be difficult, and thus we choose to make assumptions that make the processing analysis more straightforward. Encouraged by our choice of NLP model \citep{barbieri2020tweeteval}, we use two potential proxies for cyberbullying, being the hatefulness and offensiveness of textual content of tweets. By denoting tweets that meet a certain threshold as those most likely to be cyberbullying events, we can easily construct a time series to perform ensuing analyses. The provision of NLP as a filtering mechanism is an improvement that has great potential in cutting down spam and irrelevant tweets, whereas previous work only relied on the number of matching tweets based on keywords \cite{karmakar2020evaluating}. The NLP model in question is available freely for use on HuggingFace~\footnote{\url{https://huggingface.co/cardiffnlp/twitter-roberta-base-offensive}}~\footnote{\url{https://huggingface.co/cardiffnlp/twitter-roberta-base-hate}}.


To use these models, we work with the Python language in Google Colaboratory~\footnote{\url{https://research.google.com/colaboratory/}}, which provides a high performance cloud computing environment and greatly simplifies set-up of packages and other dependencies. Instructions for basic set-up to use HuggingFace are found on the website \footnote{\url{https://huggingface.co/course/chapter0/1?fw=pt}}. Template code for using the models is found on the model pages. While there is a text pre-processing step that is part of the template code, it does not contain the additional step of removing line-breaks that the authors of the model, \citet{barbieri2020tweeteval}, used in their own analysis, so it was added to the text pre-processing function. We also converted all text to lowercase to handle erratic capitalization and removed duplicate white space. The fitting procedure was generalized to collections of text using a for loop.

Using two of their models, we can produce the probability of tweet being offensive and the probability of a tweet being hateful. To filter out irrelevant data, we must properly select thresholds for each of these scores. Motivating this discussion, we first observe a few example Tweets (identifying information is censored), where H denotes the hatefulness score and O the offensiveness score according to the NLP model.

\begin{displayquote}[H: 0.07, O: 0.10]
@USER @USER Then Dr. Fauci and others should speak out every single day and defend the health and safety of US Citizens.
\end{displayquote}

\begin{displayquote}[H: 0.03, O: 0.04]
@USER Thanks. Now she needs to make it through intensive care tonight
\end{displayquote}

\begin{displayquote}[H: 0.62, O: 0.16]
@USER @USER @USER @USER @USER @USER IF ANY of these african migrants have ebola-can't that be spread thru water? or am I wrong? I thought ebola was spread w/fluids-any body fluids-can anyone inform me?thanks in advance.
\end{displayquote}

\begin{displayquote}[H: 0.86, O: 0.92]
I feel like we all went to school with a bitch like this and wanted to shove her down the mf stairs [URL]
\end{displayquote}

One particular aspect the model is good at is removing Tweets that are clearly not hateful or offensive, and are hence very unlikely to be cyberbullying, so filtering out Tweets with lower scores on these metrics may help retain better tweets. However, while not depicted, randomly sampling tweets with a high offensiveness/hatefulness rating returned many tweets containing African American English Vernacular, where the content of the tweet is not necessarily a cyberbullying event. This is part of a larger issue of systemic racial bias in a large swath of hate speech and abusive language datasets \citep{davidson2019racial}. Unfortunately, not much can be done about this in this context without further complicating the study, potentially going out of scope of the original intentions. At the very least, we can be confident that many tweets not indicative of cyberbullying will be removed after filtering, so the dataset will be relatively more representative even with the aforementioned issues.

\subsection{Subsetting Truly Offensive or Hateful Tweets}

\begin{figure}
    \centering
    \includegraphics[scale=0.5]{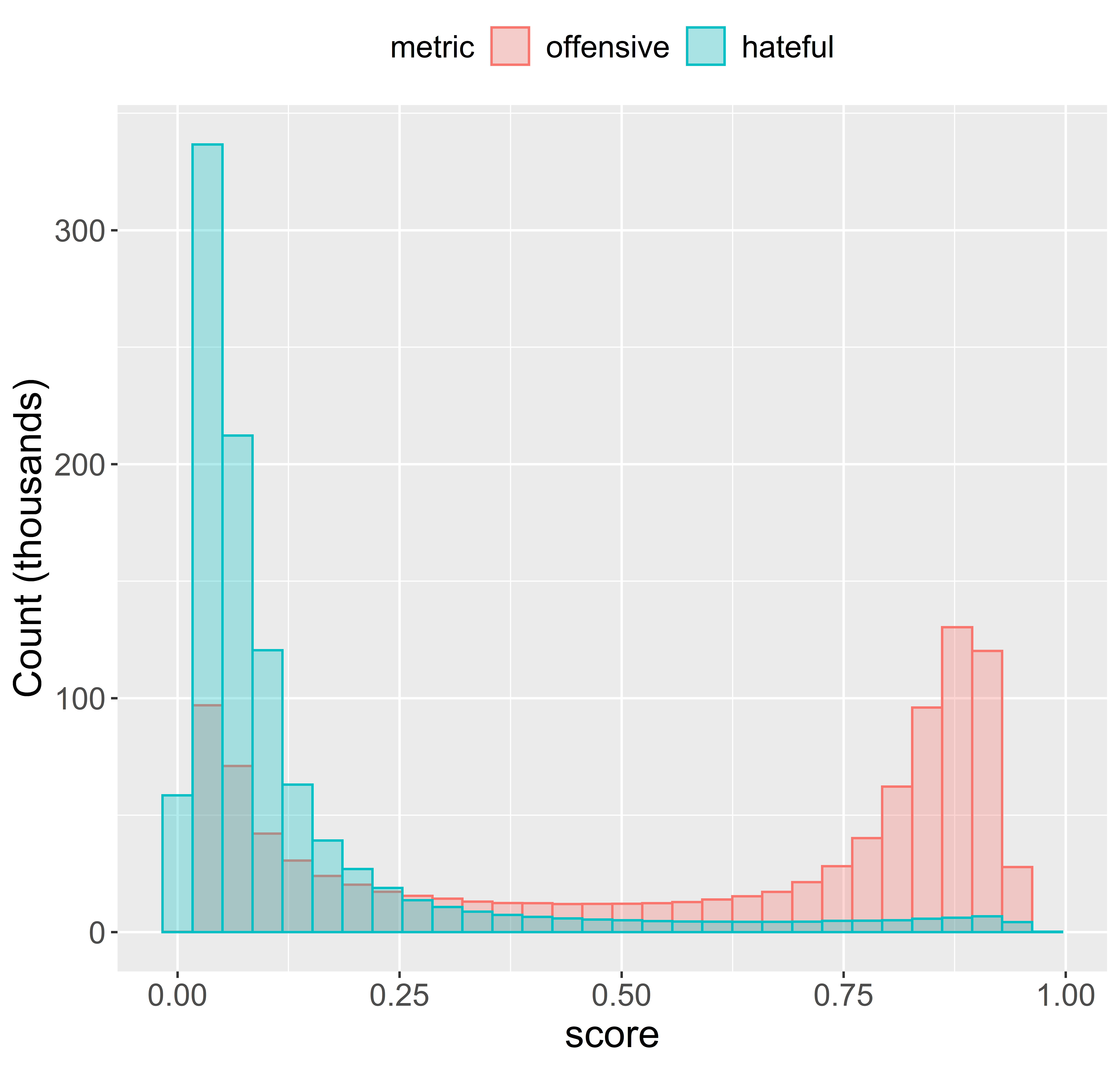}
          \caption{Modeled probabilities of offensiveness/hatefulness of tweets according to the NLP model}
    \label{fig:DistributionOfMetrics}
\end{figure}

Due to our assumption of hatefulness and offensiveness as indicators for cyberbullying, we have two different metrics to work with to help determine whether a tweet may be recorded as a cyberbullying event. We do this by selecting thresholds for each of these metrics to filter out irrelevant data. To begin, we first explore the distribution of the scores themselves. In figure \ref{fig:DistributionOfMetrics}, one may notice that the scores of offensiveness are bimodal with peaks 0 to 1, with fewer tweets near the center. The high presence of tweets near 1 is likely related to the nature of the query. In the construction of the model by \citet{barbieri2020tweeteval}, they used a dataset created by a different group for a similar task. The creators of said dataset, \citet{zampieri2019semeval}, describe a tweet as offensive if "it contains any form of non-acceptable language (profanity) or a targeted offense ... This category includes insults, threads, and posts containing profane language or swear words." Thus, if our query contains many profane keywords, we are likely to see many tweets with a high predicted probability for offensiveness. 

On the other hand, hatefulness scores are mostly lower than 0.25, and much fewer tweets have scores beyond. To reiterate, the model was fine-tuned to detect hatefulness against two target groups, being women and immigrants. The lower presence of high-probability tweets can be partially explained by queried words, as they lack many words that are explicitly targeting women or immigrants, such as those used by the authors of the dataset \citep{basile2019semeval}.

There are some tweets which the model failed to fit. These tweets all contained copious amounts of emojis which caused issues with the model's tokenizer. Out of the 1,004,466 tweets in this study, only 7 failed to process. Each of these tweets occurred on a different day, making it exceedingly unlikely for them to affect analysis.

\begin{figure}[!tbp]
  \centering
  \hspace*{-7mm}
  \begin{minipage}[b]{0.5\textwidth}
  \hspace*{5mm}
    \includegraphics[scale=0.5]{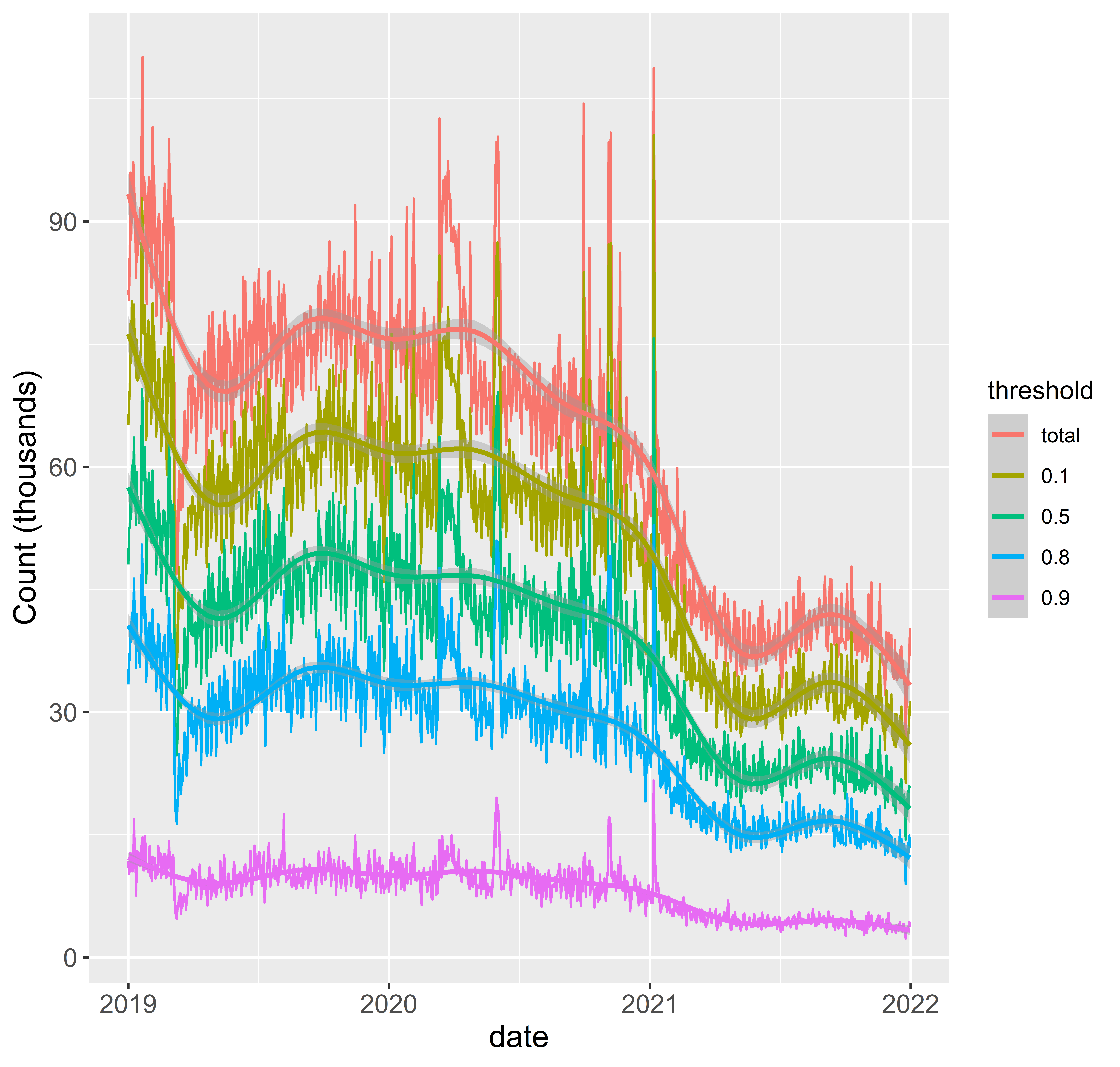}
          \caption{Results of Subsetting Method Using Offensiveness}
    \label{fig:DeltaOffensive}
  \end{minipage}
  \hfill
  \begin{minipage}[b]{0.5\textwidth}
  \hspace*{-1mm}
    \includegraphics[scale=0.5]{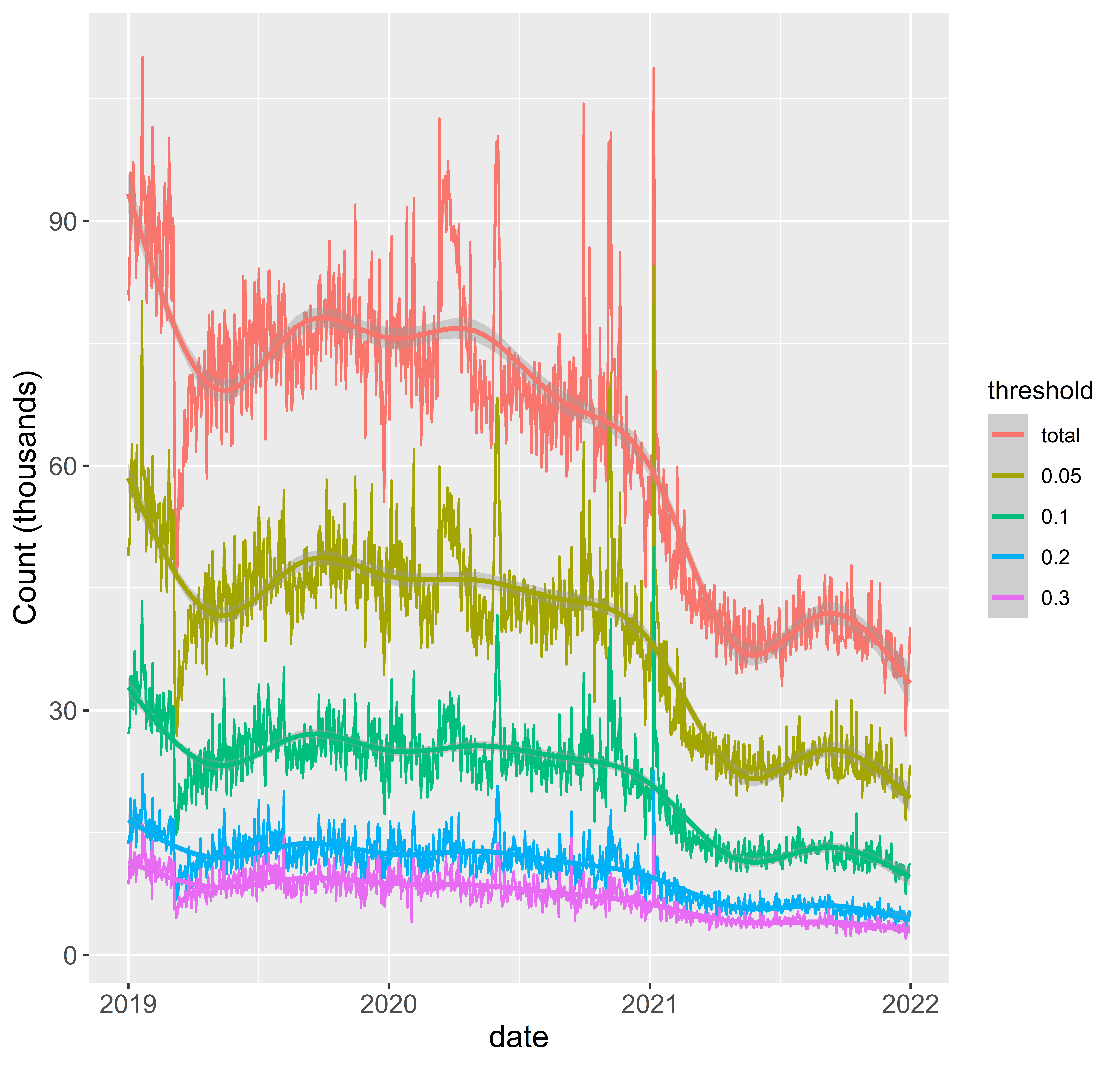}
           \caption{Results of Subsetting Method Using Hatefulness}
    \label{fig:DeltaHateful}
  \end{minipage}
\end{figure}

Now we observe the effects of subsetting tweets with scores strictly greater than a certain probability threshold, getting the percentage of those tweets from their respective daily sample, then multiplying that percent to the number reported by the count endpoint. Different thresholds are used, and the series is superimposed by a GAM smoother. The results for this process, varying offensiveness while fixing hatefulness, are displayed in figure \ref{fig:DeltaOffensive}. The local maxima are kept intact until accepting only larger scores of greater than 0.8. Likewise, for filtering on hatefulness in figure \ref{fig:DeltaHateful}, the data is much more sensitive to applying greater thresholds. When applying a threshold of 0.05 on hatefulness, the values in January 2019 were around 60,000, and then hovered around 45,000 until 2021. A similar pattern is exhibited when thresholding at the much greater value of 0.5 on offensiveness. These results are a consequence of the distribution of overall scores as shown in figure \ref{fig:DistributionOfMetrics}, with offensiveness scores being bimodal near 0 and 1, while hatefulness clusters near 0 and tapers off rapidly.


\begin{figure}[h]
    \centering
    \hspace*{-8mm}
    \includegraphics[scale=0.5]{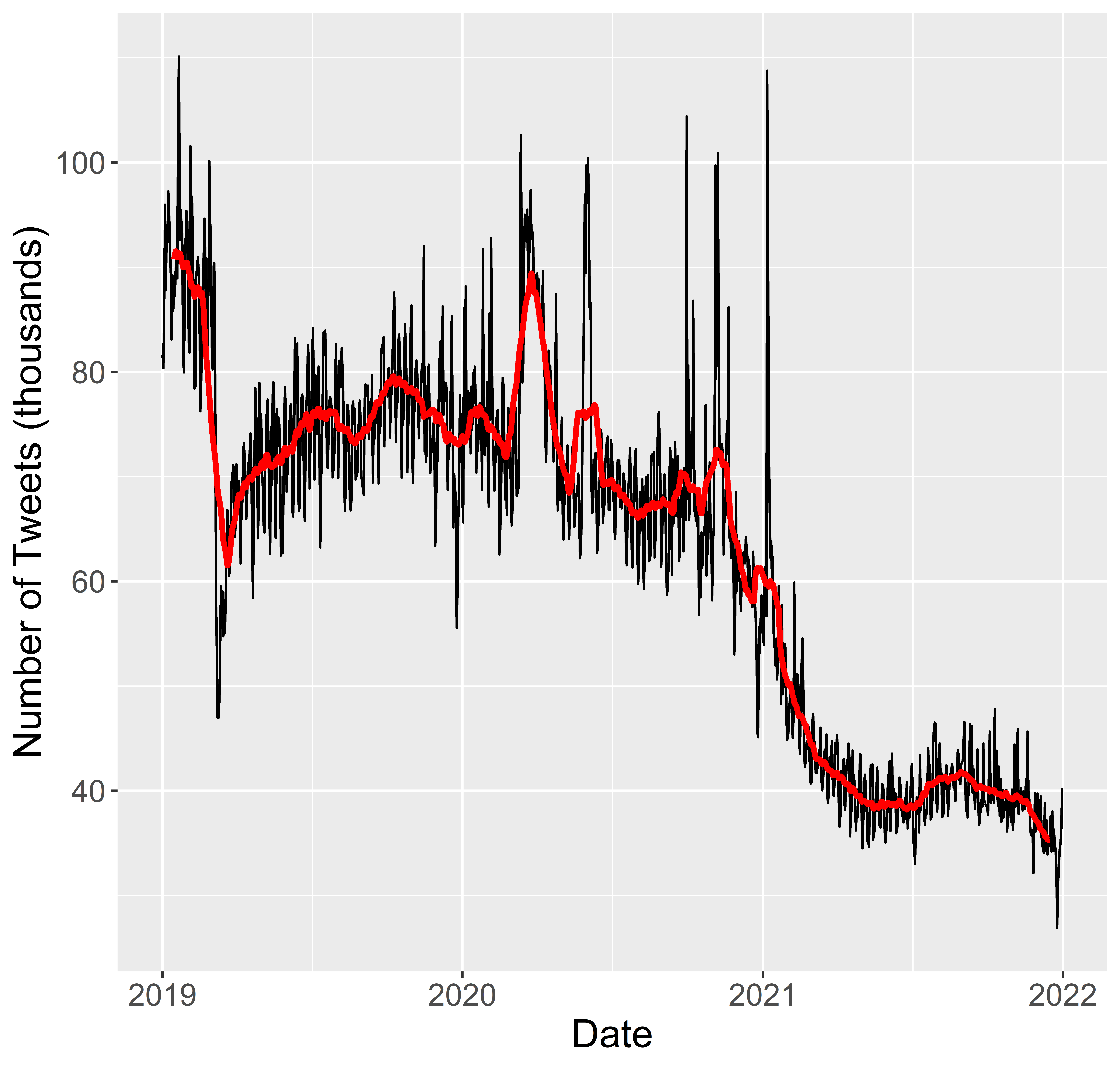}
          \caption{Daily count of total tweets containing queried keywords, 2019-2021, superimposed by a 30-day centered rolling average}
    \label{fig:CountEndpointData}
\end{figure}

Now we are tasked to choose a threshold for our dataset, then feed this data into the TVBARC model. In some problems, 0.5 may be used, but we know from reality that the occurrence of cyberbullying events is not as common as that would suggest. However, given that our query contains keywords associated with abusive behavior, it's possible that the proportion of cyberbullying events among the collected tweets will be higher. Other considerations include the NLP models' performance on their associated test data. The M-F1 score for the hatefulness model is around 50, whereas for offensiveness, it is around 80 \citep{barbieri2020tweeteval}. An additional factor is the TVBARC model itself, which may fail to converge if the data points are too large in magnitude. From above, one can observe that increasing the minimum offensiveness threshold does not greatly alter the structure of the time series, so it can be used to cut down on the scale as necessary (note that these threshold parameters can be freely tuned to a desired sensitivity). However, making the threshold too high may result in many 0s in the series, which is especially true when filtering on hatefulness. Not all offensive speech is considered cyberbullying, but hate speech can contain offensive speech like slurs, in this case directed to women or immigrants, which may be associated with cyberbullying. That being said, the data set has a non-zero number of tweets that lack offensive content but have a high probability of being hateful. Therefore, it would be wise to increase the threshold on offensiveness as necessary to allow model convergence and primarily focus on altering the hatefulness threshold. 

In this study, we will use two separate thresholds for comparison. First, some notation is established. An $x/y$ filter will refer to a filter that only accepts tweets with offensiveness probability greater than $\frac{x}{100}$ and hatefulness probability greater than $\frac{y}{100}$. In this notation, the two filters used are $25/0$ and $25/50$. Using these, of interest is rudimentary comparison with prior results before we fit to the TVBARC model. To motivate the use of this model, a visual analysis on the counts is performed.

\section{Visual Analysis on the Raw and Filtered Counts}
A visual analysis, similar to that done by \citet{karmakar2020evaluating}, allows one to deduce some trends and patterns. However, certain issues will limit the efficacy of such an analysis and justify implementation of a statistical model. First, we focus on the counts provided by the count endpoint with no thresholding, observed in figure \ref{fig:CountEndpointData}.

Starting in 2019, one sees a sudden drop in counts, which levels out until the end of 2020. In 2020, there is a prominent peak in April, which roughly agrees with the findings of \citet{karmakar2020evaluating}. Throughout the rest of 2020, there are several spikes, but they are not persistent. When 2021 begins, the counts drop yet again, but unlike 2019, the counts stay at these lower levels.

In both 2019 and 2021, one observes a decrease in counts in the beginning of the year, though this does not occur in 2020. Instead, there are many large peaks though with an overall downward trend. In 2021, the downtrend accelerates and soon levels off. It is possible that with the advent of COVID-19, the typical downward trend was disrupted as lockdowns and social isolations precipitated increased internet use \citep{candela2020impact}. The significant decrease may also be related to the findings of \citet{bacher2022covid} in their study, which uses Google search frequencies and bullying surveys to study the change in cyberbullying-related searches over time. The study shows that the log search intensity of school bullying and cyberbullying significantly decrease near the end of 2020 and beginning of 2021. But since one time series involves profanity and potentially hateful language on social media, and the other search engine data about bullying, it's possible that this is a coincidence. Additionally, due to the lack of thresholding, one cannot say that the frequency of cyberbullying events also follows the same trend. 

\begin{figure}
    \centering
    \hspace*{-5mm}
    \includegraphics[scale=0.5]{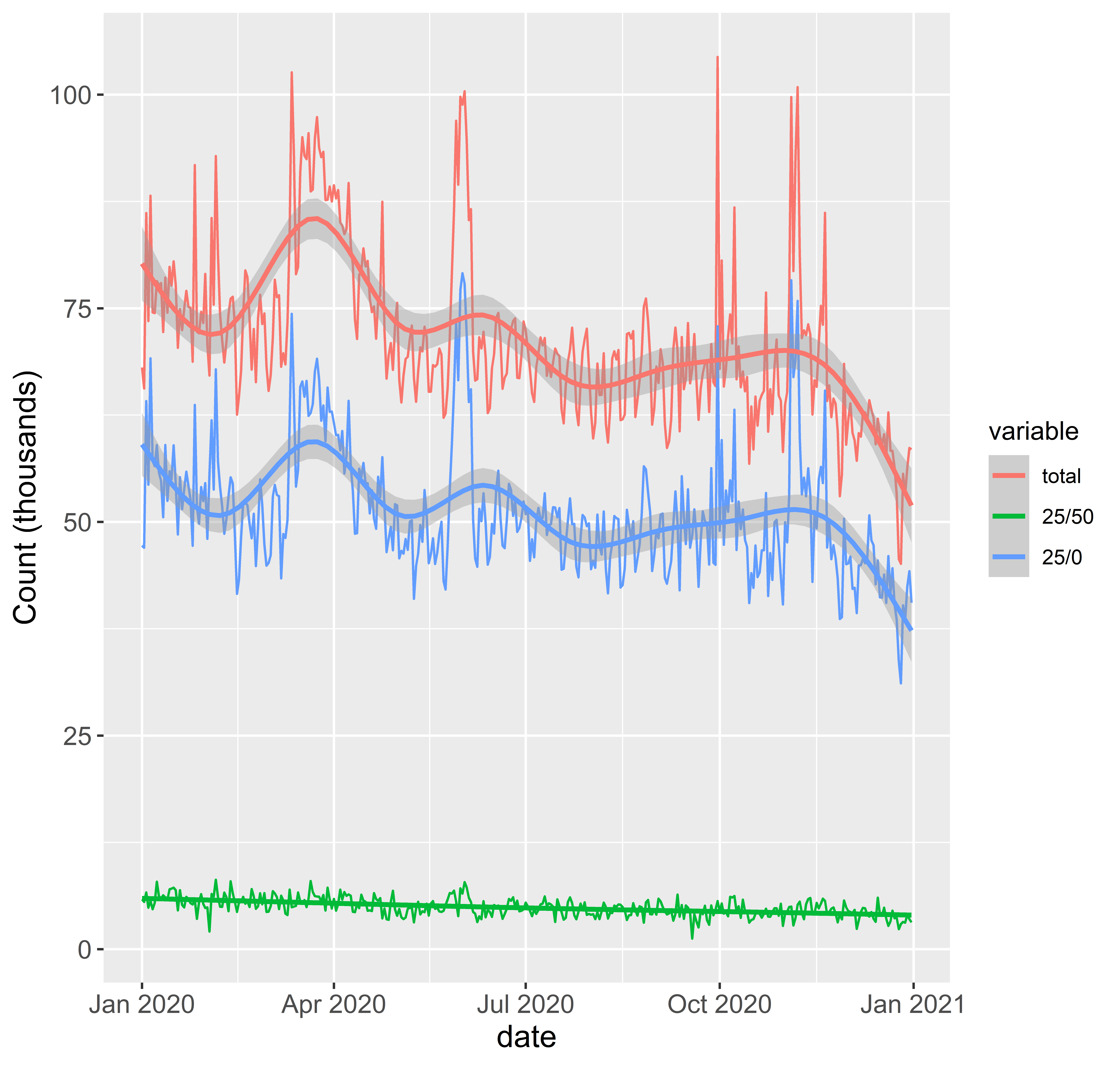}
          \caption{Total counts and filtered counts in 2020}
    \label{fig:Filtered2020Legend}
\end{figure}

We revisit the effects of thresholding, but now in the context of identifying trends. In figure \ref{fig:DeltaOffensive}, as the offensiveness threshold increases, the peaks shrink, and the trend begins to flatten. Note that it takes a threshold of 0.9 to induce some significant flattening. However, when thresholding on hatefulness, it only takes a threshold of 0.2 to achieve a similar degree of flattening. By the time we increase it to 0.5, the time series may appear constant, as shown in figure \ref{fig:Filtered2020Legend}, which shows the different threshold settings in 2020. In both the raw and 25/0 time series, the peaks are easily identifiable, and thresholding on offensiveness works to reduce the scale while preserving the peaks. When using a large hatefulness threshold such as 0.5, prominent peaks disappear.

We are not necessarily satisfied with the analysis of the raw or 25/0 data since it may not represent the actual frequency of cyberbullying events. Further, it is also clear that applying any reasonable threshold on hatefulness may cause the time series to flatten out, making it difficult to identify trends visually. Therefore, to identify trends in abusive tweets, we must employ a statistical model.

\section{Statistical Modeling and Analysis}
While there are many models we can choose from to study this data, we opt to employ a time-varying Bayesian autoregressive count (TVBARC) model based on Poisson random variables \citep{roy2020bayesian}. There are many advantages to using this form of model, such as mitigating small sample-size, accounting for dependence, modeling subtle rather than abrupt change, and only needing a single parameter. More information on the motivations of the model, including in the context of Twitter data, is available in previous works of one of the authors \citep{karmakar2020evaluating,das2020change,karmakar2021understanding}.

Suppose $Z_t$ represents the true counts of all tweets, satisfying the query, that are as hateful and offensive as our threshold dictates. Let the time series provided by the count endpoint be $Y_t$. Now let $p_t$ be the proportion of sample tweets on day $t$ that meet the threshold requirements, with $0 \leq p_t \leq 1$ for all $t$. Then, we estimate series $Z_t$ by 

$$\hat{Z_t} = X_t = p_t Y_t$$

For this model, the conditional distribution for a count time series $X_t$ given $\mathcal{F}_{t-1}=\{X_i: i\leq (t-1)\}$ is


\begin{align}
    X_t\mid \mathcal{F}_{t-1}\sim& \mathrm{Poisson}(\lambda_t)\\
    \text{ where }&\lambda_t=\mu(t/T) +
    \sum_{i=1}^p a_i(t/T) X_{t-i}
    \label{TVBARC}
\end{align}

Furthermore, $\mu(t/T)$ is the mean trend at time $t$, and $a_i(t/T)$ is the effect of the $i$th lag at time $t$. For the parameters that very over the time, a constraint on the parameter space is given as follows
\begin{align}\label{eq:parcondition}
    \mathcal{P}_1=\{\mu, a_i:\mu(x)> 0, 0\leq a_{i}(x)\leq 1, \nonumber\\  \sup_{x}\sum_{k}a_{k}(x)<1\}.
\end{align}    

In order to sample for our desired parameters, we construct a likelihood function corresponding to ~\eqref{TVBARC} given by 
\begin{align}
    L_1&\propto \exp\bigg(\sum_{t=p}^T \big[-\{\mu(t/T) + \sum_{i=1}^p a_i(t/T) X_{t-i}\big\} \nonumber \\
    &+ X_t\log \big\{\mu(t/T) + \sum_{i=1}^p a_i(t/T) X_{t-i}\}\big] \nonumber \\
    &- \sum_{j=1}^{K_1} \beta_j^2/(2c_2) - \sum_{l=0}^p \delta_l^2/(2c_1)\bigg){\mathbf 1}_{0\leq\theta_{ij}\leq 1}
\end{align}

The priors are given as
\begin{align}
\mu(x) =&\sum_{j=1}^{K_1}\exp(\beta_j)B_j(x) \label{mu_func}, \\ a_{i}(x)=&\sum_{j=1}^{K_2}\theta_{ij}M_{i}B_j(x),0\leq\theta_{ij}\leq 1 \label{coef_func},\\
    M_i=&\frac{\exp(\delta_i)}{\sum_{k=0}^p{\exp(\delta_k)}}, \quad i=1,\ldots,p,\\
    \delta_l\sim&N(0, c_1),\textrm{ for }0\leq l\leq p,\\
    \beta_{j}\sim &N(0, c_2)\textrm{ for } 1\leq j\leq K_1,\\
    \theta_{ij}\sim& U(0,1)\textrm{ for }1\leq i\leq p, 1\leq j\leq K_2\label{prior2}.
\end{align}

The $B_{j}$ terms above are B-spline basis functions and the induced prior is supported by $\mathcal{P}$. Additional details on the restrictions, as well as a verification of the support, can be found in \citet{karmakar2020evaluating}, as the very same model is used. We apply this model to the estimation $X_t$ of the true counts $Z_t$.

The fitting of the model over a count time series provides the mean trend of the series over time, as well as the coefficient values of the different autoregressive terms. From this one can deduce how the frequency of offensive/hateful tweets changed over time, as well as which lags are most related to each other. In the fitting procedure, we generate two models that differ in up to how many lags are represented. One contains up to lag 10, and the other 15. Generally, the mean trend captured is the same between these two models, though the width of the credible intervals may change. Exceptions to this will be mentioned when relevant.

\subsection{Analysis of Model Fits on Twitter Data}
We now discuss the results of the model fits for each base using different threshold or lag settings. Figures for model fits are included in the appendix due to size. The top half of a given figure shows the value of $\mu(\cdot)$, given by equation \eqref{mu_func}. The grey bands for this function are its associated 95\% credible intervals. The bottom half shows, for each lag $i$, the value of $a_i(\cdot)$ as in equation \eqref{coef_func}. When an individual lag $i$ is mentioned in the discussion, it is referring to the value of $a_i(\cdot)$.

\begin{figure*}[!h]
  \centering
  \begin{minipage}[b]{0.5\textwidth}
    \includegraphics[scale=0.65]{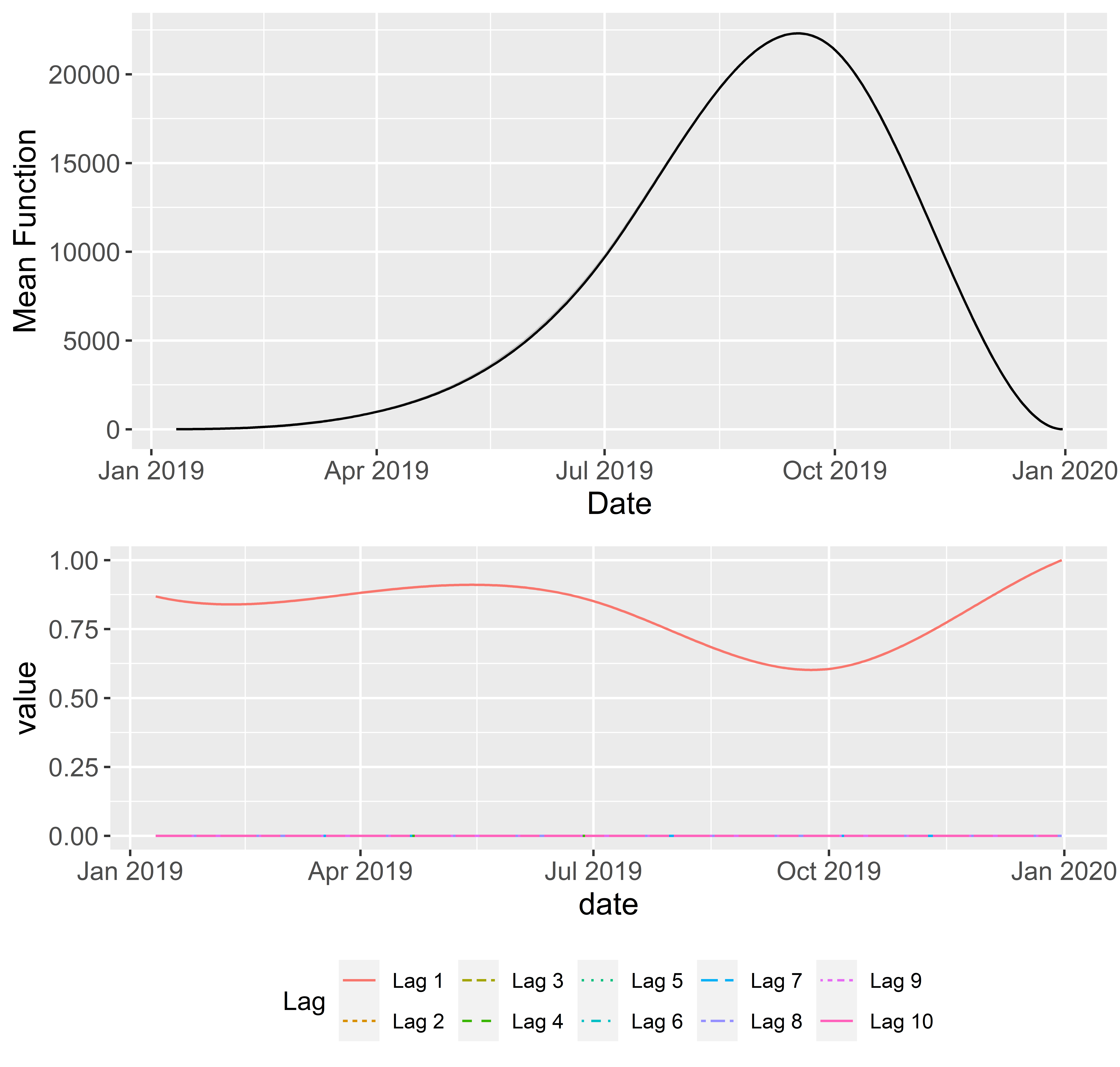}
          \caption{Mean trend and AR coefficients of Lag 10 model on 2019 Twitter data with 25/0 filter}
    \label{fig:Lag10-25/0-2019}
  \end{minipage}
  \vfill
  \begin{minipage}[b]{0.5\textwidth}
    \includegraphics[scale=0.65]{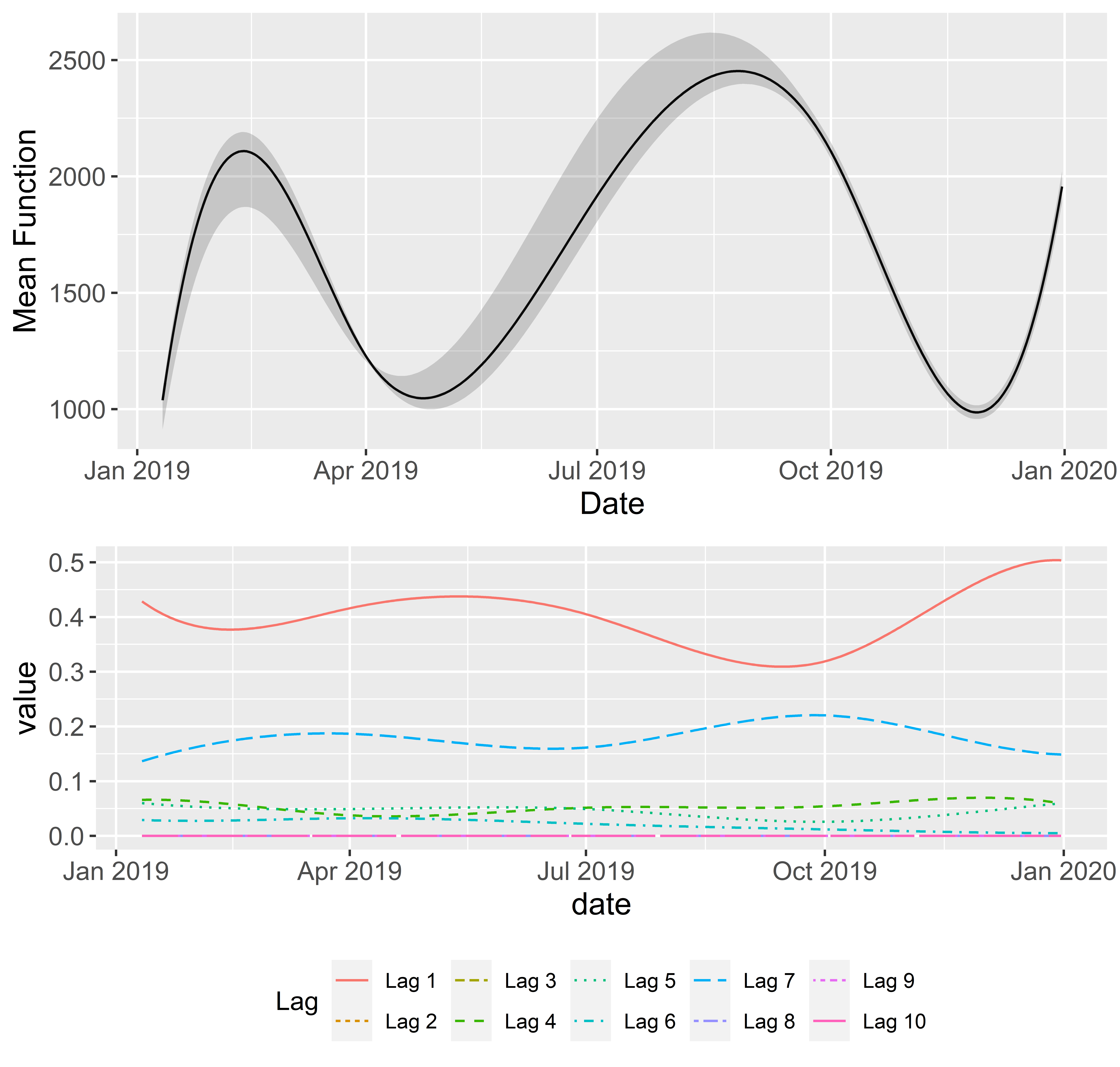}
          \caption{Mean trend and AR coefficients of Lag 10 model on 2019 Twitter data with 25/50 filter}
    \label{fig:Lag10-25/50-2019}
  \end{minipage}
\end{figure*}

Beginning with 2019, there is no significant difference between the results of the lag 10 and lag 15 model, so we only focus on lag 10. Figure \ref{fig:Lag10-25/0-2019} displays the mean trend and coefficient values of the count data in 2019 by filtering on tweets with offensiveness greater than or equal to 0.25. As shown in figure \ref{fig:DeltaOffensive}, for low thresholds, this filter reduces the scale of the original data set, preserving all structure, so this can be interpreted as a scaled down version of the original count data. Here the credible intervals are quite small, so if the true counts were to be distributed by a 10-lag process, the mean trend would look like this. However, we also see no dependence on any lag besides the first. In the analogous model fit for the 25/50 data in figure \ref{fig:Lag10-25/50-2019}, which focuses more on hateful tweets, we see a similar mean trend, capturing the peak between July and October. However, lag 7 becomes a lot more significant, and an earlier peak in the month is revealed. This process shows that the modeling procedure can uncover local extrema even when the process looks almost constant, like in figure \ref{fig:Filtered2020Legend} with the 2020 data.


The fits for 2020 are in figures \ref{fig:Lag10-25/0-2020} and \ref{fig:Lag10-25/50-2020}. The trend captured between the two threshold settings is roughly the same. However, the credible intervals for the model passing only more hateful tweets are much wider. In the 25/50 setting, lags around 7 start to get a bit more significant. One may also directly compare this to previous results. In Karmakar's previous work, the peak of cyberbullying discourse occurred in April-May 2020, while in this setting, the peak in offensive and hateful tweets occurred later in the year, between July and October \citep{karmakar2020evaluating}. A smaller peak near the beginning of the year is also observed. Compared to the previous study, the fits here have much wider credible intervals as well.


In the year of 2021, fitting the data through the 25/50 filter to the TVBARC model resulted in very different outcomes when using lag 10 (figure \ref{fig:Lag10-25/50-2021}) as opposed to 15 (figure \ref{fig:Lag15-25/50-2021}). For the 25/0 data, there is no major difference besides credible intervals and mean trend magnitude. There is an exceedingly rough similarity between the lag 10 and lag 15 model, but it is obfuscated by the very wide credible intervals in the lag 10 model. Yet in both models, lag 7 is important, as the coefficient value for this lag overtakes lag 1 during certain periods. In the lag 10 model, lag 6 is relevant, while in the lag 15 model, lag 13 is important. The results are similar, since if there is a weekly effect, one could also observe a "biweekly" relationship as well. Any dependence on lag $p$ may cause lag $p+7$ to be significant.


In general, when modeling the data on the 25/0, a scaled version of the raw counts, we most often do not see any weekly effect. However, when only considering tweets with hatefulness probability greater than 0.50, there is a very noticeable weekly effect, bringing out lags like as 5, 6, 7, 13, and 14. From this, we can infer that the frequency hateful tweets similar to those captured in the study have noticeable weekly seasonality that goes otherwise unobserved when looking at the raw counts. Additionally, the yearly pattern of two peaks in the beginning and end of the year roughly holds for 2019, 2020, and 2021. This implies that this time series also has a strong yearly seasonality. Furthermore, the mean of the hateful tweets, in most cases, follows a similar mean trend as the 25/0 series. Since this is just a scaled down version of the total data, the trend of hateful tweets mirrors that of the raw counts. This makes some sense since the filtering process takes a percentage of these counts. However, it may also reveal that the proportion of hateful tweets out of all tweets on a certain day remains about constant throughout time.

Next, correspondences with COVID-19 case counts are discussed in order to gauge possible relationships between the two time series.

\subsection{Analysis in the Context of the COVID-19 Pandemic}
\begin{figure}
    \includegraphics[scale=0.07]{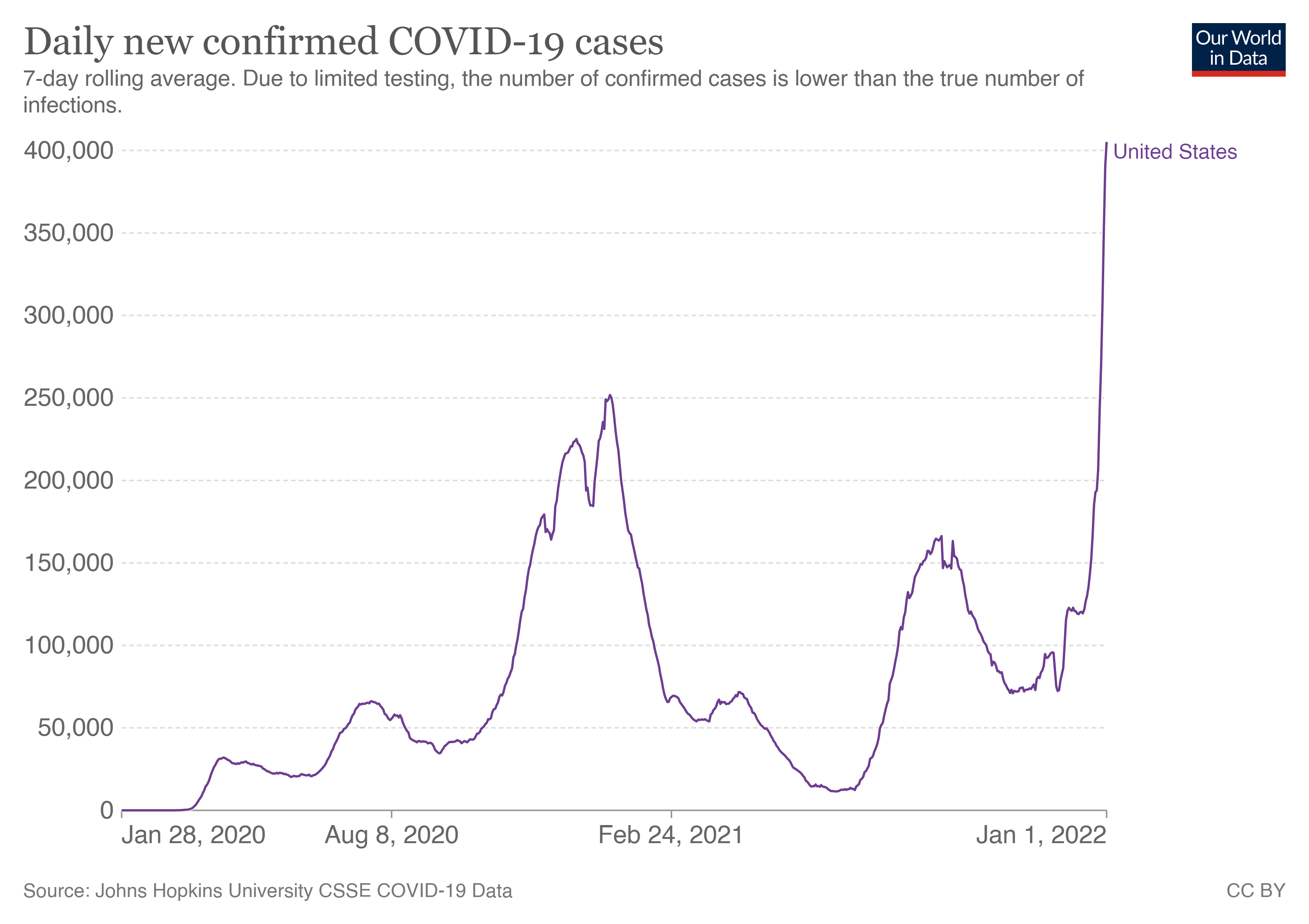}
          \caption{7-day average of confirmed COVID-19 cases in 2020-2021}
    \label{fig:COVID-19 2020-2021}
\end{figure}

By analyzing only tweets sourced from the United States, we may focus on COVID-19 counts from the same country. Depicted in figure \ref{fig:COVID-19 2020-2021} are the daily confirmed COVID-19 cases, taking a 7-day rolling average, provided by \citet{owidcoronavirus}. Beginning in 2020, there are consecutively larger peaks at around March, July, and October-December. We see a steep drop beginning near 2021, picking up during August-October. There is an exponential increase during the beginning of 2022.

As the visual analysis of the tweets in the study by \citet{karmakar2020evaluating} did not necessarily reveal all information, it is advisable to fit the COVID data into some model to better compare to the modeled Twitter data. Fortunately, since daily new cases of COVID-19 are a count time series, we can use the same model as before to reveal information about the mean trend and significant lags. Using the same data set from \citet{owidcoronavirus}, we examine daily new cases in the US each day from January 23rd, 2020, to December 31st, 2021 (since case data was not available early in the month of January 2020). We separately model the years of 2020 and 2021 and this time only use up to 10 lags. Note that because COVID-19 did not gain traction in the US in 2019, the modeled Twitter data in that year can be seen as a sort of experimental control for what the time series typically looked like pre-pandemic.

\begin{figure*}[!h]
  \centering
  \begin{minipage}[b]{0.5\textwidth}
    \includegraphics[scale=0.65]{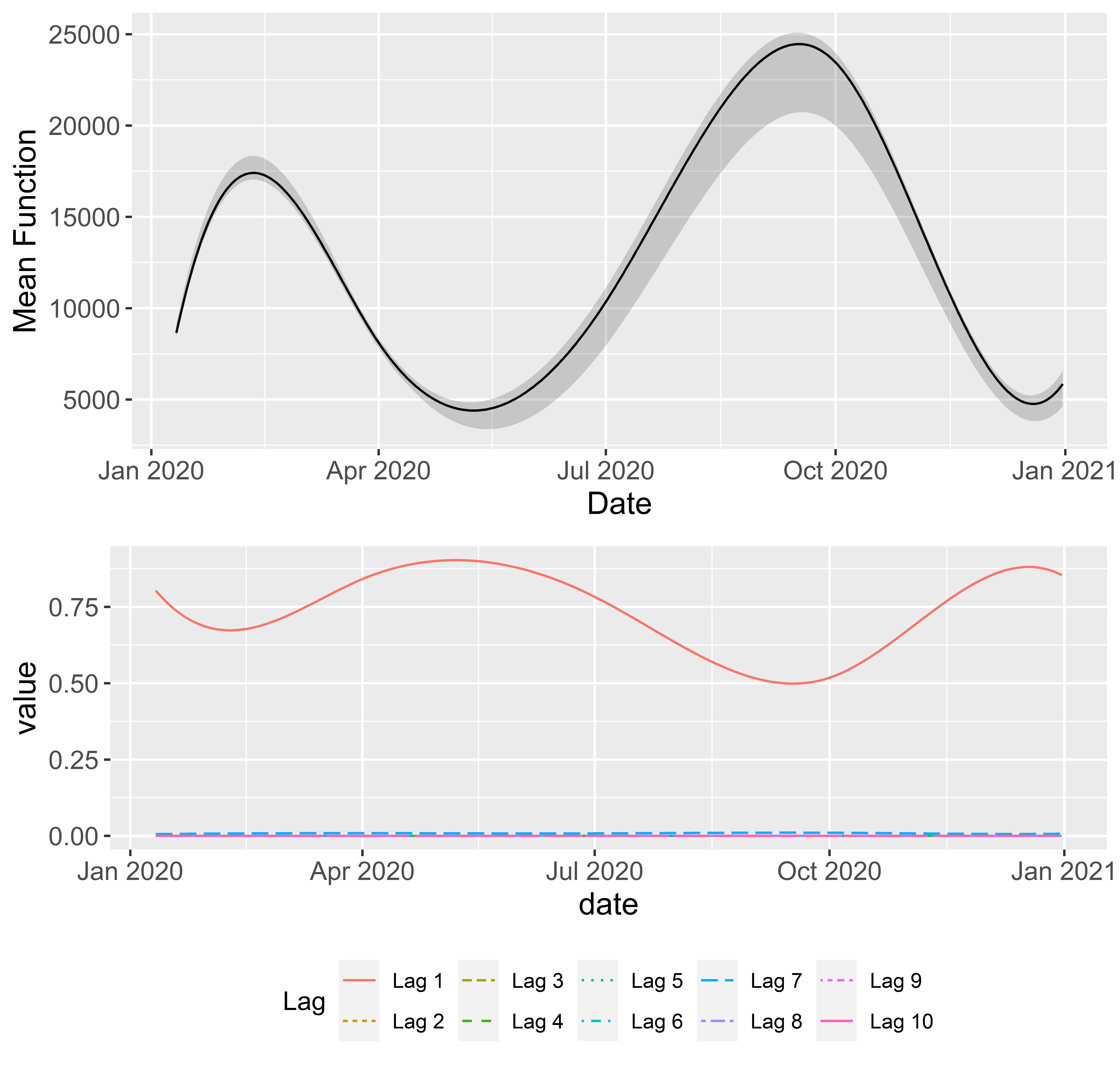}
          \caption{Mean trend and AR coefficients of Lag 10 model on 2020 Twitter data with 25/0 filter}
    \label{fig:Lag10-25/0-2020}
  \end{minipage}
  \vfill
  \begin{minipage}[b]{0.5\textwidth}
    \includegraphics[scale=0.65]{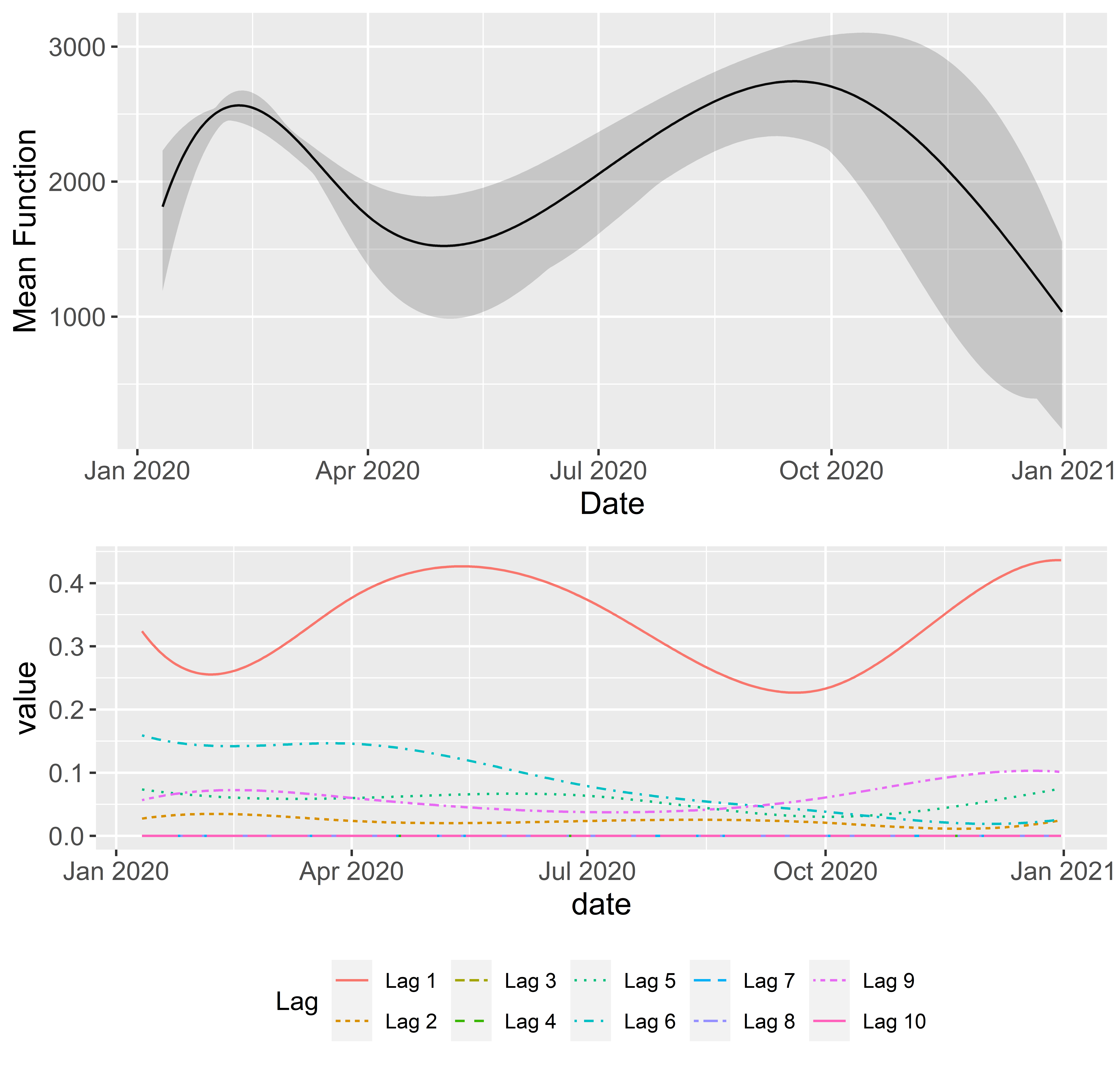}
          \caption{Mean trend and AR coefficients of Lag 10 model on 2020 Twitter data with 25/50 filter}
    \label{fig:Lag10-25/50-2020}
  \end{minipage}
\end{figure*}

\begin{figure*}[!h]
  \centering
  \begin{minipage}[b]{0.5\textwidth}
    \includegraphics[scale=0.65]{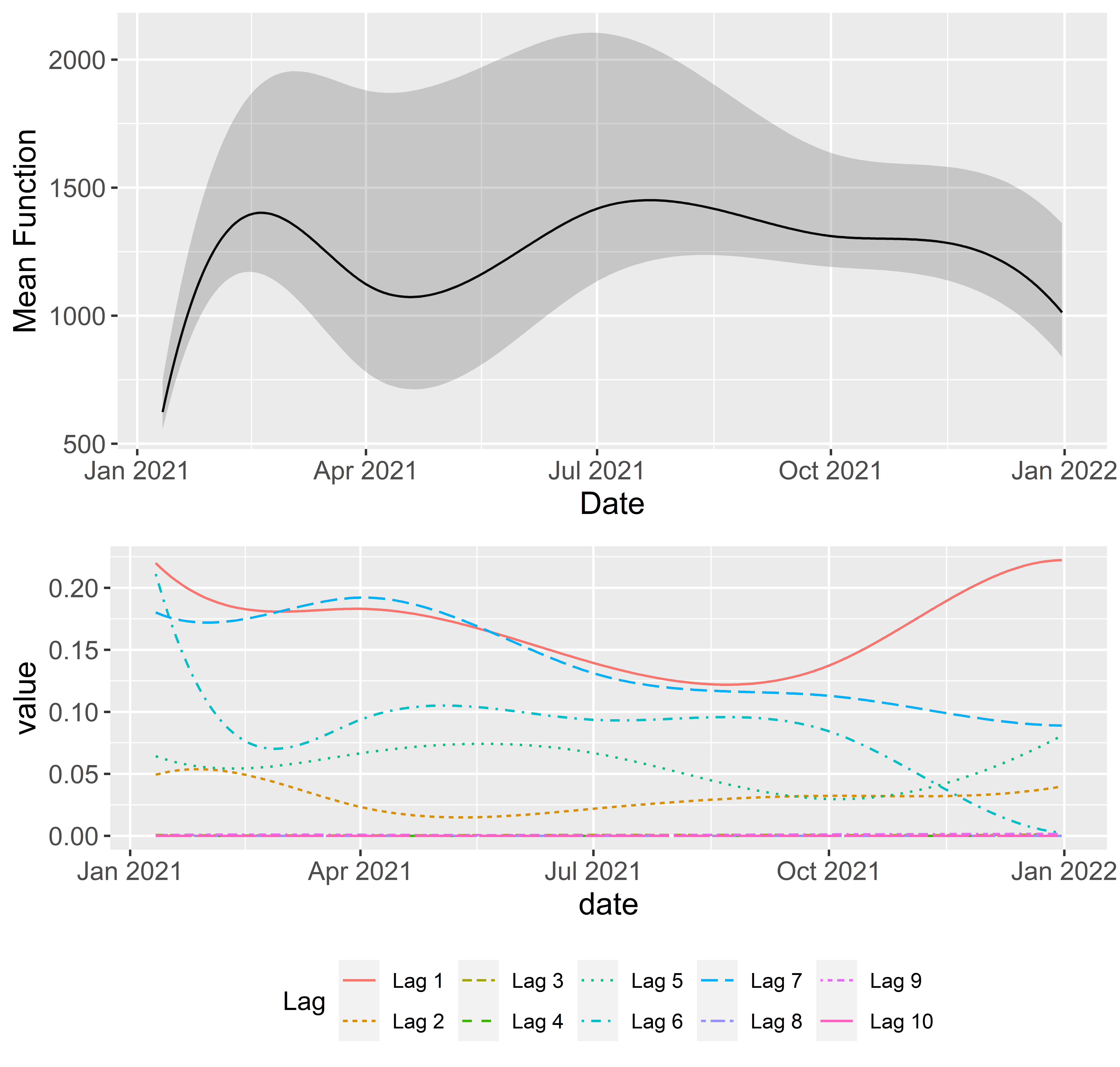}
          \caption{Mean trend and AR coefficients of Lag 10 model on 2021 Twitter data with 25/50 filter}
    \label{fig:Lag10-25/50-2021}
  \end{minipage}
  \vfill
  \begin{minipage}[b]{0.5\textwidth}
    \includegraphics[scale=0.65]{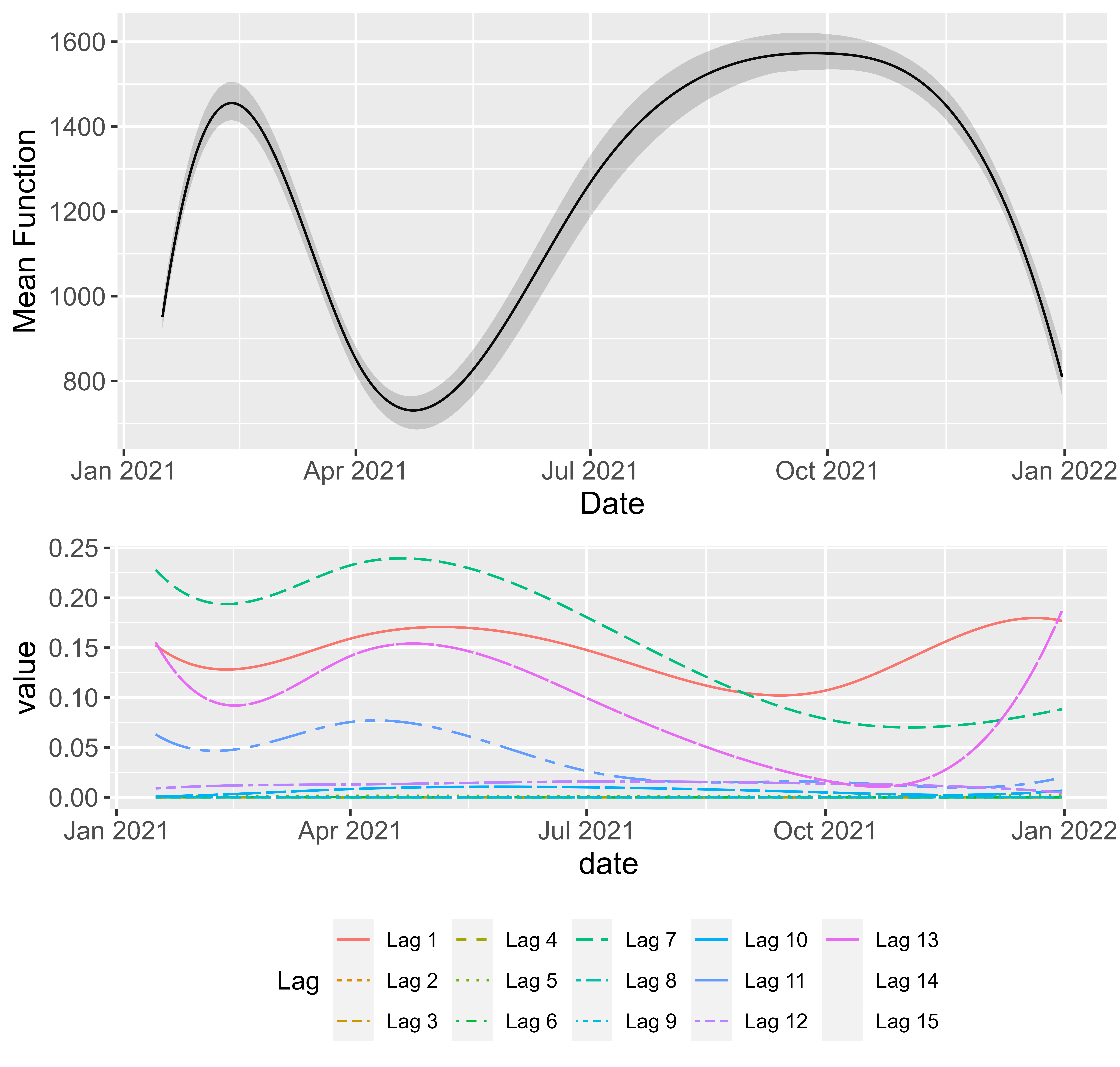}
          \caption{Mean trend and AR coefficients of Lag 15 model on 2021 Twitter data with 25/50 filter}
    \label{fig:Lag15-25/50-2021}
  \end{minipage}
\end{figure*}

Knowing the pandemic began at around 2020, we compare each year's model to see if there are any changes. Starting in 2019 with figure \ref{fig:Lag10-25/50-2019}, there is a peak in the first few months of the year, and then another during July-October. This same structure is also observed in 2020, as shown in figures \ref{fig:Lag10-25/0-2020} and \ref{fig:Lag10-25/50-2020}, as well as 2021, in figures \ref{fig:Lag10-25/50-2021} (roughly) and \ref{fig:Lag15-25/50-2021}. While the trend structure is similar, indicating the importance of seasonality, one distinction is that the secondary peak is largest in magnitude in 2020 (mean trend of 3000), followed by 2019 (about 2500) and then 2021 (about 1600). The largest peak occurring in 2020 is slightly corroborated by previous findings \citep{karmakar2020evaluating}. Further, with each year the second peak persists for longer. This may be related to a variety of things, such as increased transmission during the winter, as COVID-19 proliferation rates can change significantly with temperature \citep{mcclymont2021weather}, prompting more isolation and thus internet use, or increased cyber-aggression as a psychological response to the external stress of the pandemic \citep{wang2022covid}. 

Furthermore, a direct comparison between the mean trends can be made with each year of data to see if there are possible relationships. Consider the mean trend of tweets in figure \ref{fig:Lag10-25/50-2020} and figure \ref{fig:COVID_2020_Lag10}. There are peaks in the mean trend at around October 2020, and the lag 1 coefficient follows a very similar pattern, oscillating with a peak in the summer and a decrease in the fall. This offers some more credibility into the notion that potential cyberbullying events and COVID-19 cases increased in parallel. Now, for the year 2021 as shown in figures \ref{fig:Lag10-25/50-2021} and \ref{fig:COVID_2021_Lag10}, one can observe the wider credible interval for each of the mean trends. There is a rough correspondence in peak trend in July, but it is obfuscated by the very wide intervals. More noticeable is the fact that both datasets are most influenced by lag 1 and 7, with the Twitter data set also containing other somewhat significant lags.

For the Twitter data, the mean trend structure's similarity across these three years begs the question of whether earlier results were more of a coincidence. More specifically, previous results show that cyberbullying discourse increased roughly the same time as COVID-19 cases began to rise \citep{karmakar2020evaluating}. While these peaks occur at about the same time every year, it is important to note the change in magnitude throughout the years, dipping significantly in 2021. One may argue that the beginning of the pandemic may have brought about a sudden increase in abusive content, but as time progressed, the overall effect was to reduce such content. This angle is supported by findings using Google Trends \citep{bacher2022covid}.

\begin{figure}
    \centering
    \hspace*{-5mm}
    \includegraphics[scale=0.5]{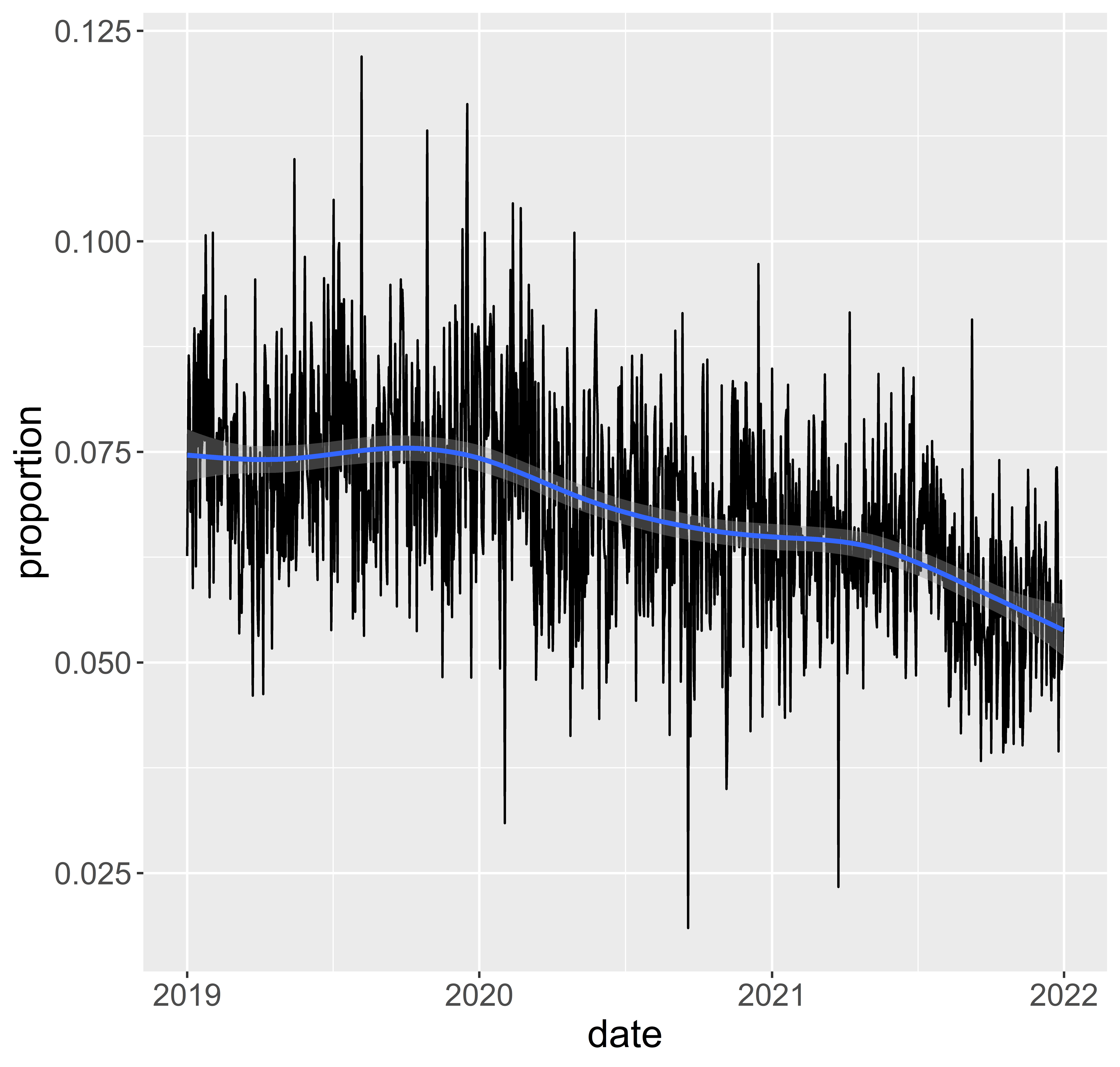}
          \caption{Proportion of tweets meeting the 25/50 threshold out of the entire daily sample superimposed by a GAM smoother}
    \label{fig:PropTrend25/50}
\end{figure}

It is also possible to study the proportions of hateful tweets, represented by $p_t$ in the model definition. These proportions give a better idea as to the relative frequency of abusive events, as the previous time series studied can be affected by the overall number of active users. Figure \ref{fig:PropTrend25/50} displays the proportion of daily tweets in the daily sample that meet the 25/50 threshold setting with a GAM smoother. The data is extremely noisy, but the smoother reveals a slight downward trend beginning just before 2020, suggesting a steady decrease in the proportion hateful content as the pandemic progressed.

\section{Discussion}
In this work, we use the official Twitter API's search endpoint to collect tweets contained keywords associated with cyberbullying over a period of 3 years. After running these tweets on an NLP model, we calculate the proportion of relevant tweets out of all tweets sampled that day using two metrics, hatefulness and offensiveness. It is our assumption that these metrics are useful in seeing whether a tweet could be considered cyberbullying or not, and that increased hateful content correlates with increased cyberbullying. The aforementioned proportion of tweets is then multiplied to the count endpoint's associated number, which is shown to be more reliable and thorough. From this we can construct a time series that is insensitive to the irregularities in daily sample sizes and better approximates the true number of hateful or offensive tweets.

\begin{figure*}[!h]
  \centering
  \begin{minipage}[b]{0.5\textwidth}
    \includegraphics[scale=0.65]{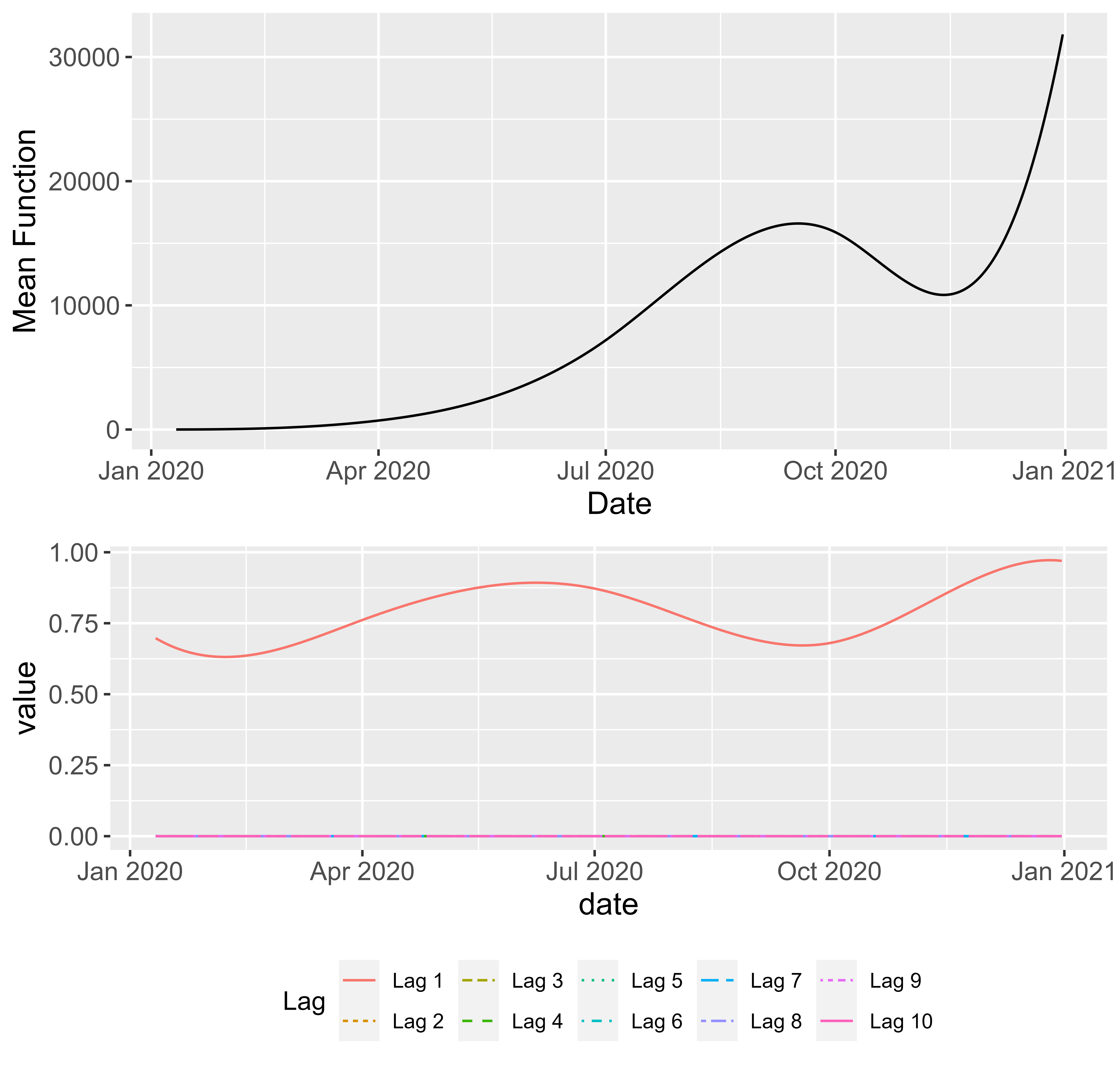}
          \caption{Mean trend and AR coefficients of Lag 10 model on 2020 COVID-19 new case data}
    \label{fig:COVID_2020_Lag10}
  \end{minipage}
  \vfill
  \begin{minipage}[b]{0.5\textwidth}
    \includegraphics[scale=0.65]{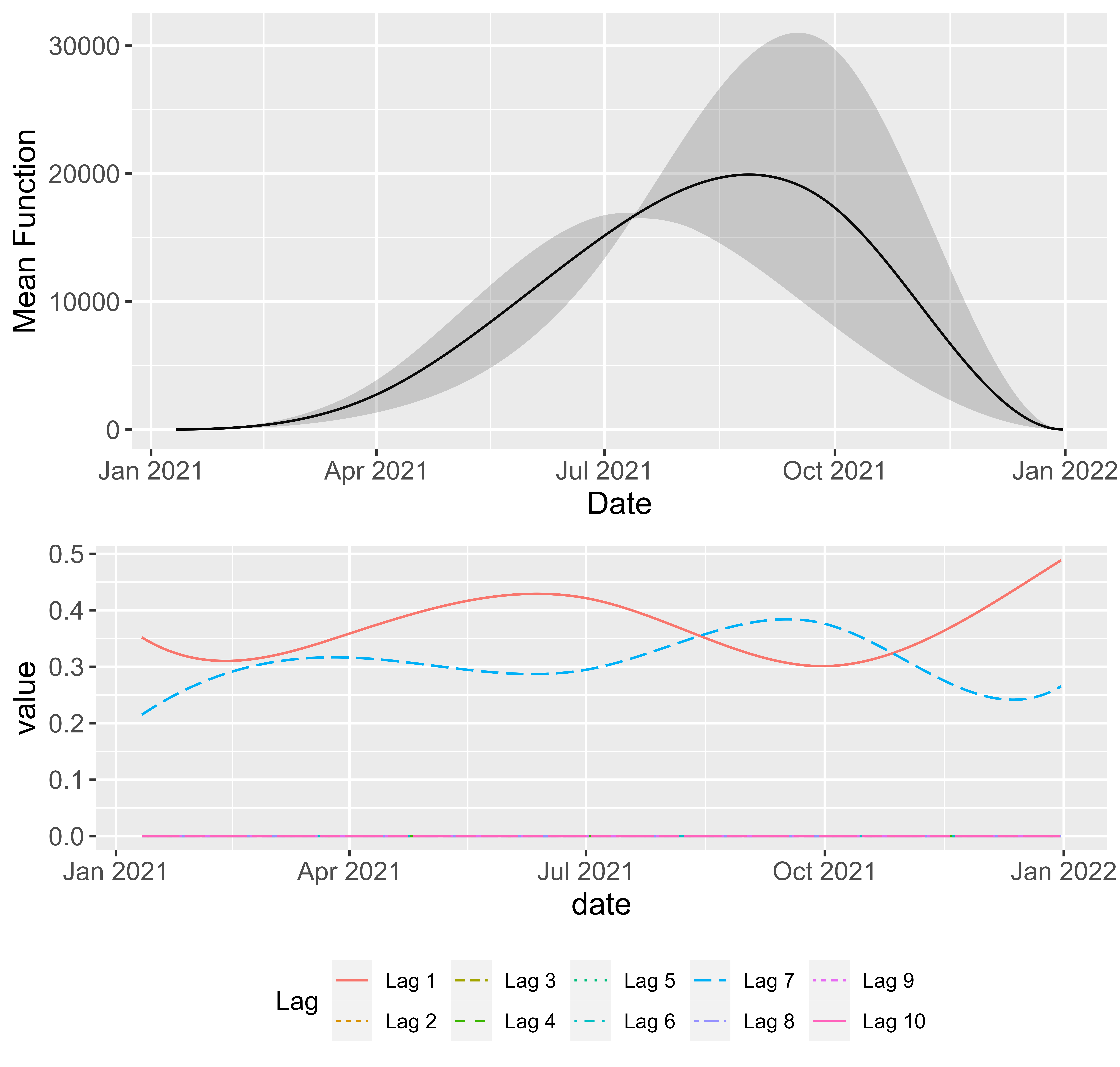}
          \caption{Mean trend and AR coefficients of Lag 10 model on 2021 COVID-19 new case data}
    \label{fig:COVID_2021_Lag10}
  \end{minipage}
\end{figure*}

The results show that the mean trend pattern of hateful tweets remains very similar through 2019, 2020, and 2021. Previous work posits that increased cyberbullying discourse happens in parallel with larger case counts \citep{karmakar2020evaluating}, which can be seen in our own results with the parallel increase in mean trend in figures \ref{fig:Lag10-25/50-2020} and \ref{fig:COVID_2020_Lag10}. However, we also notice that these peaks in the Twitter data occur in similar time frames in each of these years, including 2019 when the pandemic was not in full force. Other work suggests that cyberbullying may have been disrupted by COVID-19, as the log search intensity on Google of cyberbullying and bullying terms decreases in 2021 \citep{bacher2022covid}. Our work shows that the findings of \citet{karmakar2020evaluating} may have been a result of lack of data and suggests that hateful content may have decreased overall, agreeing with \citet{bacher2022covid}. One must take these correspondences with a grain of salt, due to the important differences between social media and search engine data \citep{li2021comparing}.

While the mean trend may have remained roughly similar, there are slight differences across the years. The secondary peak during the latter half of the year grows increasingly wide every year, which may be a consequence of virus proliferation in winter months and lifting of non-pharmaceutical interventions like lockdowns and mask mandates \citep{mcclymont2021weather, singh2021impacts}. However, these findings may be a consequence of more active Twitter users, rather than a true increase in abusive content.

Additionally, from the important seasonal lags revealed in the model fit's AR coefficients, the seasonal effect is likely the most significant factor in determining the quantity of potential cyberbullying events. Further, both the trend of Twitter data and daily case counts in 2021 (figures \ref{fig:Lag10-25/50-2021}, \ref{fig:COVID_2021_Lag10}), are both heavily influenced by lags 1 and 7.

 To summarize the possible effect of the pandemic, it is observed that the mean trend of tweets suddenly grows with the introduction of the pandemic but then cuts down significantly a year into it. Along this line, there is an increase in volume of potential cyberbullying tweets from 2019 to 2020, while there is a decrease from 2020 to 2021. This can be attributed to the raw number of hateful tweets decreasing as shown in figure \ref{fig:CountEndpointData}, while the proportion of hateful tweets out of a given day's sample also decreases across the same time frame (figure \ref{fig:PropTrend25/50}). Both the Twitter and COVID-19 case time series are influenced by seasonal lags, indicating a weekly effect.

\subsection{Limitations and Suggestions for Future Research}
A major limitation of this study is the uncertainty of how much hateful and offensive content correlates with cyberbullying events. This problem is exacerbated by racial biases in many training data sets online, the effects of which being noticeable in our own data. Further, it must be stressed that data of this nature in general cannot imply causal relationships between COVID-19 case counts and cyberbullying trends. As an example, note the finding of decreased hate speech quantity as well as proportion over the course of recent years. COVID-19 cases may not be a direct cause of this, and one may have to address several sources of confounding, such as increase internet usage precipitated by lockdown measures \citep{candela2020impact}.

The study's immediate results can only be used to infer about the approximate state of cyberbullying events in the United States among English-speaking Twitter users. One also cannot be certain that the selection of keywords produces a good sample to pull out potential cyberbullying events from. While pulling geo-tagged tweets helps in collecting a more thorough sample \citep{morstatter2013sample}, Twitter's sampling algorithm remains unknown and produces unique issues for time series analysis, such as whether it maintains a fixed sampling rate to maintain structural relationships in the count series. Additionally, the NLP model used \citep{barbieri2020tweeteval} is not necessarily state-of-the-art and was employed for its availability and ample documentation, rather than seeking the best performing algorithm known, incorporating more complex information such as social media structure and user information \citep{cheng2019xbully, dadvar2013improving}. Thus, the model's predictions on what it considers offensive or hateful may not be the most accurate. Further, the hate speech the NLP model is concerned about is primarily against women and immigrants as opposed to a broader scope of hate speech including, for instance, racism and homophobia. And again, the assumption that hateful tweets correlate with cyberbullying is not thoroughly justified.

Additionally, due to the importance of the weekly and yearly seasonality in the studied time series, employing a seasonal model in future work would better reflect the dynamics of the data. One may also consider integrating spatial methods as well, where the data set is broken down into several regions where the analysis is performed independently, like the work done by \citet{babvey2021using}. This spatial analysis can be augmented with a similar daily count time series for a continuous analysis across several regions. It is also possible to use methods of Vector Auto-regression (VAR) to model several time series simultaneously, such as COVID-19 cases, abusive tweets, and search engine data. A plethora of COVID-19 related time series from \citet{owidcoronavirus} may be used for this purpose, including data on vaccinations, ICU admissions, and deaths. As an example, studies using VAR methods show the possibility of predicting suicides using search engine data \citep{taira2021predicting} and COVID-19 cases with a great variety of variables \citep{wang2021vector}. A potentially challenging but rewarding avenue would be being able to maintain the ability to construct a count time series while employing more complex prediction models such as those in \citet{cheng2019xbully, dadvar2013improving}.


\section{Data Availability Statement}
The search endpoint data with modeled probabilities, count endpoint data, and the query used for the API requests are made available at Harvard Dataverse~\footnote{\url{https://dataverse.harvard.edu/privateurl.xhtml?token=8b866f9e-3a0c-4897-a839-b0fe6a0c2fc8}}.

\section{Funding}
The second author's research is supported by NSF DMS 2124222.

\section{Author Contribution}
C.P. collected the data, performed the analysis, and wrote the manuscript text. S.K. provided code for the Bayesian model and offered guidance on manuscript structure and communication of results. All authors reviewed the manuscript.

\section{Conflict of Interest}
The authors report no conflict of interest.

\section{Acknowledgements}
We thank the Twitter team for granting us an academic research license to carry out this study. 

\bibliography{Bullyeicc}

\begin{thebibliography}{46}
\providecommand{\natexlab}[1]{#1}
\providecommand{\url}[1]{\texttt{#1}}
\expandafter\ifx\csname urlstyle\endcsname\relax
  \providecommand{\doi}[1]{doi: #1}\else
  \providecommand{\doi}{doi: \begingroup \urlstyle{rm}\Url}\fi

\bibitem[Smith et~al.(2008)Smith, Mahdavi, Carvalho, Fisher, Russell, and
  Tippett]{smith2008cyberbullying}
Peter~K Smith, Jess Mahdavi, Manuel Carvalho, Sonja Fisher, Shanette Russell,
  and Neil Tippett.
\newblock Cyberbullying: Its nature and impact in secondary school pupils.
\newblock \emph{Journal of child psychology and psychiatry}, 49\penalty0
  (4):\penalty0 376--385, 2008.

\bibitem[Barlett(2017)]{barlett2017theory}
Christopher~P Barlett.
\newblock From theory to practice: Cyberbullying theory and its application to
  intervention.
\newblock \emph{Computers in Human Behavior}, 72:\penalty0 269--275, 2017.

\bibitem[Bonanno and Hymel(2013)]{bonanno2013cyber}
Rina~A Bonanno and Shelley Hymel.
\newblock Cyber bullying and internalizing difficulties: Above and beyond the
  impact of traditional forms of bullying.
\newblock \emph{Journal of youth and adolescence}, 42\penalty0 (5):\penalty0
  685--697, 2013.

\bibitem[Aboujaoude et~al.(2015)Aboujaoude, Savage, Starcevic, and
  Salame]{aboujaoude2015cyberbullying}
Elias Aboujaoude, Matthew~W Savage, Vladan Starcevic, and Wael~O Salame.
\newblock Cyberbullying: Review of an old problem gone viral.
\newblock \emph{Journal of adolescent health}, 57\penalty0 (1):\penalty0
  10--18, 2015.

\bibitem[Kowalski et~al.(2014)Kowalski, Giumetti, Schroeder, and
  Lattanner]{kowalski2014bullying}
Robin~M Kowalski, Gary~W Giumetti, Amber~N Schroeder, and Micah~R Lattanner.
\newblock Bullying in the digital age: A critical review and meta-analysis of
  cyberbullying research among youth.
\newblock \emph{Psychological bulletin}, 140\penalty0 (4):\penalty0 1073, 2014.

\bibitem[Kwan et~al.(2020)Kwan, Dickson, Richardson, MacDowall, Burchett,
  Stansfield, Brunton, Sutcliffe, and Thomas]{kwan2020cyberbullying}
Irene Kwan, Kelly Dickson, Michelle Richardson, Wendy MacDowall, Helen
  Burchett, Claire Stansfield, Ginny Brunton, Katy Sutcliffe, and James Thomas.
\newblock Cyberbullying and children and young people's mental health: a
  systematic map of systematic reviews.
\newblock \emph{Cyberpsychology, Behavior, and Social Networking}, 23\penalty0
  (2):\penalty0 72--82, 2020.

\bibitem[Olweus and Limber(2018)]{olweus2018some}
Dan Olweus and Susan~P Limber.
\newblock Some problems with cyberbullying research.
\newblock \emph{Current opinion in psychology}, 19:\penalty0 139--143, 2018.

\bibitem[Talevi et~al.(2020)Talevi, Socci, Carai, Carnaghi, Faleri, Trebbi,
  di~Bernardo, Capelli, and Pacitti]{talevi2020mental}
Dalila Talevi, Valentina Socci, Margherita Carai, Giulia Carnaghi, Serena
  Faleri, Edoardo Trebbi, Arianna di~Bernardo, Francesco Capelli, and Francesca
  Pacitti.
\newblock Mental health outcomes of the covid-19 pandemic.
\newblock \emph{Rivista di psichiatria}, 55\penalty0 (3):\penalty0 137--144,
  2020.

\bibitem[Wang et~al.(2021{\natexlab{a}})Wang, Shi, Que, Lu, Liu, Lu, Xu, Liu,
  Sun, Meng, et~al.]{wang2021impact}
Yunhe Wang, Le~Shi, Jianyu Que, Qingdong Lu, Lin Liu, Zhengan Lu, Yingying Xu,
  Jiajia Liu, Yankun Sun, Shiqiu Meng, et~al.
\newblock The impact of quarantine on mental health status among general
  population in china during the covid-19 pandemic.
\newblock \emph{Molecular psychiatry}, pages 1--10, 2021{\natexlab{a}}.

\bibitem[Barlett et~al.(2021{\natexlab{a}})Barlett, Simmers, Roth, and
  Gentile]{barlett2021comparing}
Christopher~P Barlett, Matthew~M Simmers, Brendan Roth, and Douglas Gentile.
\newblock Comparing cyberbullying prevalence and process before and during the
  covid-19 pandemic.
\newblock \emph{The Journal of Social Psychology}, 161\penalty0 (4):\penalty0
  408--418, 2021{\natexlab{a}}.

\bibitem[Barlett et~al.(2021{\natexlab{b}})Barlett, Rinker, and
  Roth]{barlett2021cyberbullying}
Christopher~P Barlett, Alexis Rinker, and Brendan Roth.
\newblock Cyberbullying perpetration in the covid-19 era: An application of
  general strain theory.
\newblock \emph{The Journal of Social Psychology}, 161\penalty0 (4):\penalty0
  466--476, 2021{\natexlab{b}}.

\bibitem[Bozkurt et~al.(2020)Bozkurt, Jung, Xiao, Vladimirschi, Schuwer,
  Egorov, Lambert, Al-Freih, Pete, Olcott~Jr, et~al.]{bozkurt2020global}
Aras Bozkurt, Insung Jung, Junhong Xiao, Viviane Vladimirschi, Robert Schuwer,
  Gennady Egorov, Sarah Lambert, Maha Al-Freih, Judith Pete, Don Olcott~Jr,
  et~al.
\newblock A global outlook to the interruption of education due to covid-19
  pandemic: Navigating in a time of uncertainty and crisis.
\newblock \emph{Asian Journal of Distance Education}, 15\penalty0 (1):\penalty0
  1--126, 2020.

\bibitem[Mota et~al.(2021)Mota, Silva, Costa, Aguiar, Marques, and
  Monaquezi]{mota2021mental}
Daniela Cristina~Belchior Mota, Yury Vasconcellos~da Silva, Tha{\'\i}s
  Aparecida~Ferreira Costa, Magna Helena da~Cunha Aguiar, Maria Eduarda de~Melo
  Marques, and Ricardo~Manes Monaquezi.
\newblock Mental health and internet use by university students: coping
  strategies in the context of covid-19.
\newblock \emph{Ci{\^e}ncia \& Sa{\'u}de Coletiva}, 26:\penalty0 2159--2170,
  2021.

\bibitem[Jain et~al.(2020)Jain, Gupta, Satam, and Panda]{jain2020has}
Ojasvi Jain, Muskan Gupta, Sidh Satam, and Siba Panda.
\newblock Has the covid-19 pandemic affected the susceptibility to
  cyberbullying in india?
\newblock \emph{Computers in Human Behavior Reports}, 2:\penalty0 100029, 2020.

\bibitem[Cornell et~al.(2012)Cornell, Klein, Konold, and
  Huang]{cornell2012effects}
Dewey Cornell, Jennifer Klein, Tim Konold, and Francis Huang.
\newblock Effects of validity screening items on adolescent survey data.
\newblock \emph{Psychological assessment}, 24\penalty0 (1):\penalty0 21, 2012.

\bibitem[Signorini et~al.(2011)Signorini, Segre, and
  Polgreen]{signorini2011use}
Alessio Signorini, Alberto~Maria Segre, and Philip~M Polgreen.
\newblock The use of twitter to track levels of disease activity and public
  concern in the us during the influenza a h1n1 pandemic.
\newblock \emph{PloS one}, 6\penalty0 (5):\penalty0 e19467, 2011.

\bibitem[Tumasjan et~al.(2010)Tumasjan, Sprenger, Sandner, and
  Welpe]{tumasjan2010predicting}
Andranik Tumasjan, Timm Sprenger, Philipp Sandner, and Isabell Welpe.
\newblock Predicting elections with twitter: What 140 characters reveal about
  political sentiment.
\newblock In \emph{Proceedings of the International AAAI Conference on Web and
  Social Media}, volume~4, 2010.

\bibitem[Gayo-Avello et~al.(2011)Gayo-Avello, Metaxas, and
  Mustafaraj]{gayo2011limits}
Daniel Gayo-Avello, Panagiotis Metaxas, and Eni Mustafaraj.
\newblock Limits of electoral predictions using twitter.
\newblock In \emph{Proceedings of the International AAAI Conference on Web and
  Social Media}, volume~5, pages 490--493, 2011.

\bibitem[Morstatter et~al.(2013)Morstatter, Pfeffer, Liu, and
  Carley]{morstatter2013sample}
Fred Morstatter, J{\"u}rgen Pfeffer, Huan Liu, and Kathleen Carley.
\newblock Is the sample good enough? comparing data from twitter's streaming
  api with twitter's firehose.
\newblock In \emph{Proceedings of the International AAAI Conference on Web and
  Social Media}, volume~7, pages 400--408, 2013.

\bibitem[McHugh et~al.(2019)McHugh, Saperstein, and Gold]{mchugh2019omg}
Meaghan~C McHugh, Sandra~L Saperstein, and Robert~S Gold.
\newblock Omg u\# cyberbully! an exploration of public discourse about
  cyberbullying on twitter.
\newblock \emph{Health Education \& Behavior}, 46\penalty0 (1):\penalty0
  97--105, 2019.

\bibitem[Babvey et~al.(2021)Babvey, Capela, Cappa, Lipizzi, Petrowski, and
  Ramirez-Marquez]{babvey2021using}
Pouria Babvey, Fernanda Capela, Claudia Cappa, Carlo Lipizzi, Nicole Petrowski,
  and Jose Ramirez-Marquez.
\newblock Using social media data for assessing children’s exposure to
  violence during the covid-19 pandemic.
\newblock \emph{Child Abuse \& Neglect}, 116:\penalty0 104747, 2021.

\bibitem[Bacher-Hicks et~al.(2022)Bacher-Hicks, Goodman, Green, and
  Holt]{bacher2022covid}
Andrew Bacher-Hicks, Joshua Goodman, Jennifer~G Green, and Melissa Holt.
\newblock The covid-19 pandemic disrupted both school bullying and
  cyberbullying.
\newblock Technical report, National Bureau of Economic Research, 2022.

\bibitem[Karmakar and Das(2020)]{karmakar2020evaluating}
Sayar Karmakar and Sanchari Das.
\newblock Evaluating the impact of covid-19 on cyberbullying through bayesian
  trend analysis.
\newblock In \emph{Proceedings of the European Interdisciplinary Cybersecurity
  Conference}, pages 1--6, 2020.

\bibitem[Cheng et~al.(2019{\natexlab{a}})Cheng, Li, Silva, Hall, and
  Liu]{cheng2019xbully}
Lu~Cheng, Jundong Li, Yasin~N Silva, Deborah~L Hall, and Huan Liu.
\newblock Xbully: Cyberbullying detection within a multi-modal context.
\newblock In \emph{Proceedings of the twelfth acm international conference on
  web search and data mining}, pages 339--347, 2019{\natexlab{a}}.

\bibitem[Dadvar et~al.(2013)Dadvar, Trieschnigg, Ordelman, and
  De~Jong]{dadvar2013improving}
Maral Dadvar, Dolf Trieschnigg, Roeland Ordelman, and Franciska De~Jong.
\newblock Improving cyberbullying detection with user context.
\newblock In \emph{Advances in Information Retrieval: 35th European Conference
  on IR Research, ECIR 2013, Moscow, Russia, March 24-27, 2013. Proceedings
  35}, pages 693--696. Springer, 2013.

\bibitem[Cheng et~al.(2019{\natexlab{b}})Cheng, Guo, Silva, Hall, and
  Liu]{cheng2019hierarchical}
Lu~Cheng, Ruocheng Guo, Yasin Silva, Deborah Hall, and Huan Liu.
\newblock Hierarchical attention networks for cyberbullying detection on the
  instagram social network.
\newblock In \emph{Proceedings of the 2019 SIAM international conference on
  data mining}, pages 235--243. SIAM, 2019{\natexlab{b}}.

\bibitem[Huang et~al.(2014)Huang, Singh, and Atrey]{huang2014cyber}
Qianjia Huang, Vivek~Kumar Singh, and Pradeep~Kumar Atrey.
\newblock Cyber bullying detection using social and textual analysis.
\newblock In \emph{Proceedings of the 3rd International Workshop on
  Socially-aware Multimedia}, pages 3--6, 2014.

\bibitem[Wiegand et~al.(2019)Wiegand, Ruppenhofer, Schmidt, and
  Greenberg]{wiegand2019inducing}
Michael Wiegand, Josef Ruppenhofer, Anna Schmidt, and Clayton Greenberg.
\newblock Inducing a lexicon of abusive words--a feature-based approach.
\newblock In \emph{Proceedings of the 2018 Conference of the North American
  Chapter of the Association for Computational Linguistics: Human Language
  Technologies, June 1-June 6, 2018, New Orleans, Louisiana, Volume 1 (Long
  Papers)}, pages 1046--1056. Association for Computational Linguistics, 2019.

\bibitem[Nand et~al.(2016)Nand, Perera, and Kasture]{nand2016bullying}
Parma Nand, Rivindu Perera, and Abhijeet Kasture.
\newblock “how bullying is this message?”: A psychometric thermometer for
  bullying.
\newblock In \emph{Proceedings of COLING 2016, the 26th International
  Conference on Computational Linguistics: Technical Papers}, pages 695--706,
  2016.

\bibitem[Cortis and Handschuh(2015)]{cortis2015analysis}
Keith Cortis and Siegfried Handschuh.
\newblock Analysis of cyberbullying tweets in trending world events.
\newblock In \emph{Proceedings of the 15th International Conference on
  Knowledge Technologies and Data-driven Business}, pages 1--8, 2015.

\bibitem[Barbieri et~al.(2020)Barbieri, Camacho-Collados, Espinosa-Anke, and
  Neves]{barbieri2020tweeteval}
Francesco Barbieri, Jose Camacho-Collados, Luis Espinosa-Anke, and Leonardo
  Neves.
\newblock {TweetEval:Unified Benchmark and Comparative Evaluation for Tweet
  Classification}.
\newblock In \emph{Proceedings of Findings of EMNLP}, 2020.

\bibitem[Li et~al.(2021)Li, Goodell, and Shen]{li2021comparing}
Yue Li, John~W Goodell, and Dehua Shen.
\newblock Comparing search-engine and social-media attentions in finance
  research: Evidence from cryptocurrencies.
\newblock \emph{International Review of Economics \& Finance}, 75:\penalty0
  723--746, 2021.

\bibitem[Thelwall(2015)]{thelwall2015evaluating}
Mike Thelwall.
\newblock Evaluating the comprehensiveness of twitter search api results: A
  four step method.
\newblock \emph{Cybermetrics: International Journal of Scientometrics,
  Informetrics and Bibliometrics}, 18-19:\penalty0 1, 2015.

\bibitem[Davidson et~al.(2019)Davidson, Bhattacharya, and
  Weber]{davidson2019racial}
Thomas Davidson, Debasmita Bhattacharya, and Ingmar Weber.
\newblock Racial bias in hate speech and abusive language detection datasets.
\newblock \emph{arXiv preprint arXiv:1905.12516}, 2019.

\bibitem[Zampieri et~al.(2019)Zampieri, Malmasi, Nakov, Rosenthal, Farra, and
  Kumar]{zampieri2019semeval}
Marcos Zampieri, Shervin Malmasi, Preslav Nakov, Sara Rosenthal, Noura Farra,
  and Ritesh Kumar.
\newblock Semeval-2019 task 6: Identifying and categorizing offensive language
  in social media (offenseval).
\newblock \emph{arXiv preprint arXiv:1903.08983}, 2019.

\bibitem[Basile et~al.(2019)Basile, Bosco, Fersini, Debora, Patti, Pardo,
  Rosso, Sanguinetti, et~al.]{basile2019semeval}
Valerio Basile, Cristina Bosco, Elisabetta Fersini, Nozza Debora, Viviana
  Patti, Francisco Manuel~Rangel Pardo, Paolo Rosso, Manuela Sanguinetti,
  et~al.
\newblock Semeval-2019 task 5: Multilingual detection of hate speech against
  immigrants and women in twitter.
\newblock In \emph{13th International Workshop on Semantic Evaluation}, pages
  54--63. Association for Computational Linguistics, 2019.

\bibitem[Candela et~al.(2020)Candela, Luconi, and Vecchio]{candela2020impact}
Massimo Candela, Valerio Luconi, and Alessio Vecchio.
\newblock Impact of the covid-19 pandemic on the internet latency: A
  large-scale study.
\newblock \emph{Computer Networks}, 182:\penalty0 107495, 2020.

\bibitem[Roy and Karmakar(2020)]{roy2020bayesian}
Arkaprava Roy and Sayar Karmakar.
\newblock Bayesian semiparametric time varying model for count data to study
  the spread of the covid-19 cases.
\newblock \emph{arXiv preprint arXiv:2004.02281}, 19:\penalty0 21, 2020.

\bibitem[Das et~al.(2020)Das, Kim, and Karmakar]{das2020change}
Sanchari Das, Andrew Kim, and Sayar Karmakar.
\newblock Change-point analysis of cyberbullying-related twitter discussions
  during covid-19.
\newblock \emph{arXiv preprint arXiv:2008.13613}, 2020.

\bibitem[Karmakar and Das(2021)]{karmakar2021understanding}
Sayar Karmakar and Sanchari Das.
\newblock Understanding the rise of twitter-based cyberbullying due to covid-19
  through comprehensive statistical evaluation.
\newblock In \emph{Proceedings of the 54th Hawaii international conference on
  system sciences}, 2021.

\bibitem[Ritchie et~al.(2020)Ritchie, Mathieu, Rodés-Guirao, Appel, Giattino,
  Ortiz-Ospina, Hasell, Macdonald, Beltekian, and Roser]{owidcoronavirus}
Hannah Ritchie, Edouard Mathieu, Lucas Rodés-Guirao, Cameron Appel, Charlie
  Giattino, Esteban Ortiz-Ospina, Joe Hasell, Bobbie Macdonald, Diana
  Beltekian, and Max Roser.
\newblock Coronavirus pandemic (covid-19).
\newblock \emph{Our World in Data}, 2020.
\newblock https://ourworldindata.org/coronavirus.

\bibitem[McClymont and Hu(2021)]{mcclymont2021weather}
Hannah McClymont and Wenbiao Hu.
\newblock Weather variability and covid-19 transmission: a review of recent
  research.
\newblock \emph{International journal of environmental research and public
  health}, 18\penalty0 (2):\penalty0 396, 2021.

\bibitem[Wang et~al.(2022)Wang, Luo, Tu, Xiao, and Hu]{wang2022covid}
Qiong Wang, Xiao Luo, Ruilin Tu, Tao Xiao, and Wei Hu.
\newblock Covid-19 information overload and cyber aggression during the
  pandemic lockdown: The mediating role of depression/anxiety and the
  moderating role of confucian responsibility thinking.
\newblock \emph{International journal of environmental research and public
  health}, 19\penalty0 (3):\penalty0 1540, 2022.

\bibitem[Singh et~al.(2021)Singh, Shaikh, Hauck, and Miraldo]{singh2021impacts}
Surya Singh, Mujaheed Shaikh, Katharina Hauck, and Marisa Miraldo.
\newblock Impacts of introducing and lifting nonpharmaceutical interventions on
  covid-19 daily growth rate and compliance in the united states.
\newblock \emph{Proceedings of the National Academy of Sciences}, 118\penalty0
  (12), 2021.

\bibitem[Taira et~al.(2021)Taira, Hosokawa, Itatani, Fujita,
  et~al.]{taira2021predicting}
Kazuya Taira, Rikuya Hosokawa, Tomoya Itatani, Sumio Fujita, et~al.
\newblock Predicting the number of suicides in japan using internet search
  queries: Vector autoregression time series model.
\newblock \emph{JMIR public health and surveillance}, 7\penalty0 (12):\penalty0
  e34016, 2021.

\bibitem[Wang et~al.(2021{\natexlab{b}})Wang, Zhou, and Chen]{wang2021vector}
Qinan Wang, Yaomu Zhou, and Xiaofei Chen.
\newblock A vector autoregression prediction model for covid-19 outbreak.
\newblock \emph{arXiv preprint arXiv:2102.04843}, 2021{\natexlab{b}}.

\end{thebibliography}

\section{Appendix: Bayesian Model Fit Graphs}

\end{sloppypar}
\end{document}